\documentclass[12pt]{article}
\newlength{\abstractwidth}
\usepackage{dsfont}
\usepackage{amssymb}
\usepackage{latexsym}
\usepackage{braket}
\usepackage{graphicx}
\usepackage{subcaption}
\usepackage[utf8]{inputenc}
\usepackage[usenames]{color}

\usepackage{hyperref}
\definecolor{darkred}{rgb}{0.8,0.1,0.1}
\hypersetup{colorlinks=true, linkcolor=darkred, citecolor=blue, linktoc=page}

\renewcommand{\thanks}[1]{\footnote{#1}}

\newcommand{\bea}{\begin{eqnarray}}
\newcommand{\eea}{\end{eqnarray}}
\newcommand{\ee}{\end{equation}}
\newcommand{\be}{\begin{equation}}

\def\a{\alpha}
\def\b{\beta}
\def\g{\gamma}
\def\e{\epsilon}
\def\f{\varphi}

\def\s{\sigma}

\def\k{\kappa}

\def\cA{{\cal A}}
\def\cB{{\cal B}}
\def\cC{{\cal C}}

\def\cG{{\cal G}}

\def\cO{{\cal O}}

\def\Re{{\rm Re}}

\def\half{ {1\over 2}}
\def\p{\partial}

\def\RR{{\mathbb R}}

\def\QQ{{\mathbb Q}}

\def\no{\nonumber}

\title{(p,q)-strings probing five-brane webs}
\author{Justin Kaidi}
\date{}

\makeatletter\def\l@subsubsection#1#2{}%
\makeatother

\begin{document}
\maketitle
\begin{center}
{ \sl Mani L. Bhaumik Institute for Theoretical Physics}\\
{ \sl Department of Physics and Astronomy }\\
{\sl University of California, Los Angeles, CA 90095, USA}

{\tt \small jkaidi@physics.ucla.edu}

\end{center} 
\vspace{0.3 in}
\begin{abstract}

In recent work, globally well-defined Type IIB supergravity solutions with geometry $AdS_6 \times S^2$ warped over a Riemann surface $\Sigma$ were constructed and conjectured to describe the near-horizon geometry of $(p,q)$ five-brane webs in the conformal limit. In the present paper, we offer more evidence for this interpretation of the supergravity solutions in terms of five-brane webs. In particular, we explore the behavior of probe $(p,q)$-strings in certain families of these $AdS_6 \times S^2\times \Sigma$ backgrounds and compare this behavior to that predicted by microscopic, brane web considerations. In the microscopic picture, we argue that the embedding of a probe string may give rise to the formation of string junctions involving open strings anchored on the branes of the web. We then identify a quantity on the supergravity side that is conjectured to be equivalent to the total junction tension in a class of backgrounds corresponding to brane webs with four semi-infinite external five-branes. In the process, we will show that for general brane web backgrounds, the minimal energy probe string embeddings do not coincide with the embeddings preserving half of the background supersymmetries.

\end{abstract}

\newpage
\tableofcontents
\newpage
\section{Introduction}

In a series of recent works  \cite{DHoker:2016ujz,DHoker:2016ysh,DHoker:2017mds}, globally well-defined Type IIB supergravity solutions preserving 16 supersymmetries and with geometry $AdS_6 \times S^2$ warped over a Riemann surface $\Sigma$ were constructed. The solutions were obtained by solving the BPS equations and were given entirely in terms of a pair of locally holomorphic functions $\cA_\pm$ living on $\Sigma$. For the solutions to be regular and physical, $\Sigma$ was required to have a boundary. In the cases studied so far, $\Sigma$ has been taken to be the unit disc with or without punctures. In the above papers, it was conjectured that these supergravity solutions describe the near-horizon geometry of $(p,q)$ five-brane webs in the conformal limit, i.e. in the limit where the internal branes of the web are shrunk down to zero size. 

To actually compare quantities on the supergravity side with those on the brane web side, a dictionary of identifications between the two is necessary. In \cite{DHoker:2016ysh,DHoker:2017mds}, the following two dictionary entries were conjectured:
\begin{enumerate}
\item The poles of the function $\p \cA_+$ on $\Sigma$ are the locations of the semi-infinite external five-branes of the web.
\item  The residue of $\p \cA_+$ at a given pole encodes the $(p,q)$-charges of the semi-infinite external five-brane corresponding to that pole.
\end{enumerate}
 The partial derivative on $\cA_\pm$ in these dictionary entries is with respect to a local complex coordinate $w$ on $\Sigma$. 

As will be reviewed in detail in Section 2, the poles of $\p \cA_+$ are located along the boundary of $\Sigma$, where the spacetime $S^2$ vanishes. With just this information, two methods of testing the brane web dictionary entries present themselves. The first is to examine the behavior of the full spacetime metric as a pole is approached - if the first entry is correct, one must find that in the near-pole limit the spacetime metric reduces to the near-brane geometry of a single infinite $(p,q)$ five-brane, the precise $(p,q)$-charges of which are dictated by the second entry. The second check of the brane web interpretation is to calculate the three-form flux through a non-contractible $S^3$ surrounding a pole (the points on the boundary of $\Sigma$ where the $S^2$ vanish correspond to the north and south poles of this $S^3$) in the supergravity solutions and to compare this to the flux expected from a $(p,q)$ five-brane. In \cite{DHoker:2017mds}, the dictionary entries were shown to pass both of these tests. 

A third dictionary entry was put forward in \cite{DHoker:2017zwj}, where it was noted that when punctures are added to the unit disc $\Sigma$,  axion-dilaton monodromy is allowed in the solutions. This signals the addition of $(p,q)$ seven-branes to the web, motivating the following dictionary entry,
\begin{enumerate}
\setcounter{enumi}{2}
\item Punctures in $\Sigma$ can be identified with (so far mutually local) $(p,q)$ seven-branes in the face of the web. 
\end{enumerate}
Though this is an interesting avenue of research - especially for the realization of exceptional symmetries if configurations of mutually non-local seven-branes could be constructed \cite{Gaberdiel:1997ud} - the solutions with monodromy are significantly more complicated, and so we do not deal with them here.

This paper is based on the idea that further information about the brane web dictionary can be obtained by embedding various probes into the background geometry and observing their behavior. In the present paper, we will study the energy profiles of probe $(p,q)$-strings embedded in $AdS_6 \times S^2 \times \Sigma$ backgrounds given by the supergravity solutions, and attempt to explain the observed profiles in terms of the brane web interpretation specified by the dictionary entries above. In all cases, we will find consistency between the results on the supergravity side and the results obtained via a microscopic (though semiclassical) picture of open strings extending between the branes of the web and the probe string. In particular, we will see that the open strings anchored on branes of the web may form three-pronged string junctions with the probe string, and in many cases it is these junctions which dictate the characteristic features of the energy profiles. This may also be interpreted as non-trivial evidence for the open/closed string duality proposed in \cite{Polchinski:1995mt} and explored further (in a context similar to the current one) in \cite{Lunin:2008tf}. 

In addition to providing strong evidence for the first two dictionary entries, we will also conjecture a fourth dictionary entry, which seems to hold at least in a specific class of brane web configurations. This conjectural entry equates a certain simple quantity on the supergravity side with the total tension of the string junction formed when embedding an F1- or D1-string probe. In some very special cases, this supergravity quantity may be expressed as the norm-square of a function which is closely related to the defining supergravity functions $\cA_\pm$, evaluated on the boundary of $\Sigma$. The restriction to the boundary of $\Sigma$ here is largely for computational tractability - we will see that when restricted to $\p\Sigma$, many quantities on the supergravity side have simplifying expansions.

Along the way, an interesting conflict between minimal energy conditions and supersymmetry conditions for the probe string embedding will be observed. We find that only in a certain class of backgrounds will it be possible to have minimal energy $(p,q)$ probe string embeddings which preserve half of the supersymmetries of the background. We will provide some preliminary results towards a classification of such backgrounds, though we do not give any complete classification.

The outline of this paper is as follows. In Section 2, we offer a brief review of the supergravity solutions of \cite{DHoker:2016ujz,DHoker:2016ysh,DHoker:2017mds}, with special focus on the aspects of relevance to us in this paper. In Section 3, we explore fundamental string embeddings which preserve the maximum possible number of bosonic symmetries. These are the embeddings which have a chance of preserving half of the background symmetries, and their simplicity will allow us to make quantitative comparisons between supergravity and microscopic quantities in later sections. In the process, we will observe a conflict between string embeddings which are minimal energy and those which preserve background supersymmetries. In Section 4, we explore the analogous class of embeddings for probe D1-strings. Finally, in Section 5 we offer a series of checks of the brane web interpretation of the first two dictionary entries, as well as offer a tentative fourth dictionary entry. We conclude with a discussion of directions for future work, as well as a few words on the interpretation of the probe embeddings as BPS objects in the dual five-dimensional SCFTs.

\section{Global Supergravity Solutions}
\setcounter{equation}{0}
We begin with a brief review of the 1/2-BPS supergravity solutions obtained in \cite{DHoker:2016ujz,DHoker:2016ysh,DHoker:2017mds}. For a more thorough introduction to the information of this section, the reader is recommended to check these references directly.

The supergravity solutions of the above references were motivated largely by the desire to study five-dimensional SCFTs via holography. As such, we first recall the symmetries of such a SCFT. In five-dimensions, the superconformal algebra is the exceptional Lie superalgebra $F(4)$. More precisely, it is the particular real form of $F(4)$, often denoted $F(4;2)$, which has maximal bosonic subalgebra $SO(2,5) \oplus SO(3)$ \cite{LieAlgDictionary,Parker:1980af}. These symmetries motivate the Ansatz $AdS_6 \times S^2 \times \Sigma$ for the dual spacetime. Here $\Sigma$ is a two-dimensional Riemann surface over which the $AdS_6$ and $S^2$ can in general be warped. In particular, this Ansatz has the following (Einstein-frame) metric 
\bea
\label{met}
ds^2 = f_6^2 \,ds_{AdS_6}^2 + f_2^2\, ds_{S^2}^2 + 4 \rho^2 \,| d \omega|^2
\eea
where the metric factors $f_6^2$, $f_2^2$, and $\rho^2$ are functions only of the local complex coordinates $w, \bar w$ of $\Sigma$. To preserve the full symmetry algebra, the RR 2- and 4-form potentials must have the following form, 
\bea
C_{(2)} = \cC \, \omega_{S^2} \hspace{1.5 in} C_{(4)} = 0
\eea
with $\cC$ a function on $\Sigma$ and $ \omega_{S^2}$ the volume element of the 2-sphere. 
In \cite{DHoker:2016ujz}, the BPS equations on this Ansatz were integrated to obtain explicit results for the supergravity fields in terms of two locally holomorphic functions $\cA_\pm$ on the Riemann surface. The solutions are most easily expressed in terms of the following variables, 
\bea
\label{genericintro}
\k^2 &=& - |\p_w \cA_+|^2 + |\p_w \cA_-|^2
\no\\
\p_w \cB &=& \cA_+ \p_w \cA_- - \cA_- \p_w \cA_+ 
\no\\
\cG &=& |\cA_+|^2 - |\cA_-|^2 + \cB + \bar \cB 
\no\\
R + {1 \over R} &=& 2 + 6 {\k^2 \cG \over |\p_w \cG|^2}
\eea
In terms of these variables and an integration constant $c_6$, the metric factors are found to be
 \bea
 \label{improvedmetricfactors}
 f_6^2 = c_6^2 \sqrt{6 \cG} \left({1 + R \over 1 - R} \right)^{\half} \hspace{0.5 in}  f_2^2 ={ c_6^2 \over 9} \sqrt{6 \cG} \left({1 - R \over 1 + R} \right)^{3 \over 2} \hspace{0.5 in} \rho^2 = {\k^2 \over \sqrt{6 \cG}} \left({1 + R \over 1 - R} \right)^{\half} \hspace{0.2 in}
 \eea
and the axion-dilaton field $B$, defined as 
\bea
\label{axdil}
B = {1 + i \tau \over 1 - i \tau} \hspace{2 in} \tau = \chi + i e^{-2 \phi}
\eea
is given in terms of the holomorphic data by  
\bea
\label{axdil2}
 B &=& {\p_w \cA_+ \p_{\bar w} \cG - R\, \p_{\bar w} \bar \cA_- \p_w \cG \over R \,\p_{\bar w} \bar \cA_+ \p_w \cG - \p_w \cA_- \p_{\bar w} \cG}
 \eea
 Finally, the flux potential $\cC$ is \footnote{Note that in \cite{DHoker:2016ysh}, there is a factor of $R$ missing from the second term in the numerator.}
 \bea
 \cC = {4 i c_6^2 \over 9} \left[ {\p_{\bar w} \bar \cA_- (R^2 + 1) \p_w \cG - 2 R\, \p_w \cA_+ \p_{\bar w} \cG \over \k^2 \, (R+1)^2} - \bar \cA_- - 2 \cA_+\right]
 \eea
 To have physically reasonable solutions, we must further impose a set of reality, positivity, and regularity conditions, which constrain $\Sigma$ to be a Riemann surface with boundary. Furthermore, these conditions require that
  \bea
 \k^2 > 0 \hspace{1 in} \cG > 0
 \eea
 in the interior of $\Sigma$, as well as
 \bea
  \k^2 \,|_{\p \Sigma} =  \cG\, |_{\p \Sigma} = 0
 \eea
 on the boundary. These requirements imply that $R = 1$ on $\p \Sigma$, which in turn implies that $f_2^2$ vanishes on $\p \Sigma$. Hence along the boundary of $\Sigma$ the spacetime $S^2$ vanishes, and so the boundary of $\Sigma$ is not a boundary of the spacetime. 

Given a particular Riemann surface $\Sigma$, one may obtain an explicit expression for the locally holomorphic functions $\cA_\pm$. For the rest of this paper, we will take $\Sigma$ to be either the unit disc with no punctures or (equivalently) the upper-half plane with a point at infinity. While working with the unit disc has some conceptual advantages, calculations are far simpler in the upper-half plane. Thus unless otherwise specified, we take $\Sigma$ to be the upper-half plane in all formulas. In this case, one finds
\bea
\label{cA}
\cA_\pm = \cA^0_\pm + \sum_{\ell = 1}^L Z^\ell_\pm \log(w - r_\ell)
\eea
with the coefficients $Z^\ell_\pm$ satisfying $Z^\ell_\pm = - \bar {Z^\ell_\mp}$. The $\cA^0_\pm$ are complex constants determined by the $L-1$ regularity conditions
\bea
 \label{regularity1}
 \cA^0 Z_-^k + \bar \cA^0 Z_+^k  + \sum_{\ell \neq k} Z^{[\ell\,k]} \log|r_\ell - r_k| = 0
 \eea 
For compactness, we have defined $Z^{[\ell\,k]} = Z_+^\ell Z_-^k - Z_+^k Z_-^\ell$. From the above, each choice of $1\leq k \leq L-1$ gives a new condition that must be obeyed by regular supergravity solutions. The $L$-th condition is redundant due to ``charge conservation" $\sum_k Z_+^k =0$, which can be shown to be a property of the solutions.  

The $r_\ell$ in (\ref{cA}) and (\ref{regularity1})  are $L$ points on the boundary of $\Sigma$. We will often refer to these points as ``poles", since they are poles of the functions $\p \cA_\pm$. In the brane-web interpretation of the solutions, these poles are the locations of the semi-infinite external $(p,q)$ five-branes of the web along the boundary of $\Sigma$. Furthermore, the $(p,q)$-charges of the $k$-th external brane are given in terms of the coefficients $Z_+^k$ by 
\bea
\label{pqcharges}
p_k = {8 \over 3} \,c_6^2\, \mathrm{Re}\left(Z_+^k \right) \hspace{1 in} q_k =- {8 \over 3} \,c_6^2\, \mathrm{Im}\left(Z_+^k \right)
\eea
These identifications make up the first two dictionary entries mentioned in the Introduction. 

As was also noted in the Introduction, to support the brane-web interpretation we may explore the behavior of the solutions in the near-pole limit. To do so, we begin by switching to the string frame via $\tilde{f}_6^2 = e^\phi f_6^2$, and likewise for $\tilde{f}_2^2$ and $\tilde{\rho}^2$. In terms of polar coordinates $r, \theta$ centered around the $k$-th pole $r_k$, the near-pole limit corresponds to the $r \ll 1$ regime. In this regime, some works shows that
\bea
\label{metricnearpole}
\tilde{f}_6^2 \approx 2\, c_6^2\, |Z_+^k - Z_-^k| \, |\log r| \hspace{0.7 in} \tilde{f}_2^2 \approx {2 \over 3}\, c_6^2 \,|Z_+^k - Z_-^k|  \sin^2 \theta\hspace{0.7 in} \tilde{\rho}^2 \approx {1 \over 6} |Z_+^k - Z_-^k| r^{-2}
\no
\eea
Putting these together in the string frame metric, we obtain 
\bea
\label{sfmnp}
d\tilde{s}^2 \approx {2 \over 3} |Z_+^k - Z_-^k| \left(3\, c_6^2 |\log r| ds^2_{AdS_6} + {dr^2 \over r^2} + d\theta^2 + c_6^2 \, \sin^2\theta \,ds^2_{S^2} \right)
\eea 
The last two terms combine to give a smooth $S^3$ without conical defect only when $c_6^2 = 1$. We thus set $c_6^2$ to one for the rest of this paper. The radius of curvature of the $AdS_6$ diverges as $r \rightarrow 0$, so in the near-pole limit the $AdS_6$ becomes six-dimensional Minkowski space, 
\bea
\label{nearpolemetric}
d\tilde{s}^2 \approx ds^2_{\RR^{1,5}} +   {2 \over 3}\, |Z_+^k - Z_-^k| \left({dr^2 \over r^2} +ds^2_{S^3} \right)
\eea
Under the identifications (\ref{pqcharges}) this indeed reproduces the known result for the near-horizon geometry of an infinite $(p,q)$ five-brane \cite{Lu:1998vh}. 

Before moving on, it will be useful to know the behavior of various supergravity functions under overall scalings of the background charges, i.e. under scalings of all charges by $(p_i , q_i) \rightarrow \lambda (p_i, q_i)$. The reader may easily verify that under such scalings we have
\bea
\label{overallscale}
\cA_\pm &\rightarrow& \lambda \cA_\pm \hspace{1.3 in} R\, \,\rightarrow\, \,R
 \no\\
\k^2 &\rightarrow& \lambda^2 \k^2 \hspace{1.27 in} f_6^2\,\, \rightarrow\, \,|\lambda| f_6^2
 \no\\
\cG &\rightarrow& \lambda^2 \cG \hspace{1.35 in} B\, \,\rightarrow \,\,B
\eea

\section{F1-String Embeddings}
\setcounter{equation}{0}

We now turn towards the embedding of probe F1-strings in the above geometries. We will be particularly interested in probe string embeddings which preserve half of the background supersymmetries, i.e. 8 of 16 background supersymmetries.\footnote{These embeddings are 1/2-BPS on the background, but only 1/4-BPS in Type IIB, since the background itself breaks half of the Type IIB supersymmetries. Nevertheless, we will refer to these embeddings as simply ``$1/2$-BPS" in what follows.} Motivated by this, we begin by focusing on probe string embeddings that preserve the maximum possible amount of bosonic symmetries - we will make some comments in the Discussion on relaxing this restriction. In particular, if we take our probe string to wrap an $AdS_2$ in $AdS_6$, it will break the $SO(5,2)$ isometry algebra to an $SO(2,1)$ and a transverse $SO(4)$. The maximum amount of symmetry that can be preserved is then $SO(2,1)\oplus SO(4) \oplus SO(3)$ - the last factor is the symmetry algebra of the spacetime $S^2$. 

Indeed, this is a good set of symmetries to have. As stated in the previous section, the complex form of the superconformal algebra in five dimensions is $F(4)$. This has several sub-superalgebras, one of which is the (complex) algebra $A_1 \oplus D(2,1;2)$ \cite{LieAlgDictionary,Parker:1980af}. This is relevant to our current considerations for two reasons. First, $A_1 \oplus D(2,1;2)$ contains (real) subalgebras $SL(2, \RR) \oplus SU(2) \oplus SU(2) \oplus SU(2)$, which are equivalent to the symmetry algebras in the last paragraph via 
\bea
SO(2,1) \cong  SL(2, \RR) \hspace{0.5 in} SO(4) \cong SU(2) \oplus SU(2) \hspace{0.5 in}  SO(3) \cong SU(2)_R
\no
\eea
Second, the factor $D(2,1;2)$ is a superalgebra containing 8 supercharges, which (if unbroken) is the desired amount of supersymmetry for a 1/2-BPS probe string embedding. Thus we see that when the probe string is embedded in such a way as to preserve the maximum amount of bosonic symmetries, there are just enough bosonic symmetries to support the full $A_1 \oplus D(2,1;2)$, and hence the desired 1/2-BPS configuration.\footnote{Embeddings which break the $A_1$ while leaving the $D(2,1;2)$ intact can also give rise to 1/2-BPS configurations, and will be mentioned in the Discussion.}

In light of these symmetries, it is natural for us to choose the following parameterization for the $AdS_6$ metric,
\bea
ds_{AdS_6}^2 = \cosh^2u \, ds_{AdS_2}^2 + \sinh^2u \, ds_{S^3}^2 + du^2 
\eea
where the $AdS_2$ factor is
\bea
ds_{AdS_2}^2 = {1 \over \sinh^2 y_1} \left(dy_1^2 - d y_0^2 \right) 
\eea
with $u, y_0$ taking values on the real line and $y_1 \in \RR_+$. 

To preserve the $SO(4)$ isometry, the location of the probe string along $u$ should be such that the $S^3$ vanishes, lest we single out a particular point on the $S^3$ at which to embed the string. Hence we take the probe to be located at $u = 0$. Likewise, to preserve the $SO(3)$ symmetry, we must locate the string somewhere along the boundary of $\Sigma$, where the $S^2$ vanishes. To preserve the $SO(2,1)$ isometry, the entire string worldsheet should be located at just a single point on $\p \Sigma$. However, exactly where along the boundary this point should be located is not dictated by symmetries. Rather, it should be determined by requiring that the embedding extremizes the string action, as well as that the embedding be 1/2-BPS. Surprisingly, we will find that in a large number of backgrounds we cannot simultaneously satisfy these two criteria. That is, the locations $x$ on $\p \Sigma$ which give extrema of the action are not generally locations where supersymmetry can be preserved. We will see this in more detail in Section 3.3.

Before that though, note that the action of a fundamental string wrapping $AdS_2$ is given by the usual Nambu-Goto action,
\bea
\label{action}
S = - {1 \over 2 \pi \ell_s^2} \int d^2 \sigma \sqrt{-\mathrm{det}\,\p_\alpha X^i \, \p_\beta X^j \, \tilde{G}_{i j} } = - { \tilde{f}_6^2 \over 2 \pi \ell_s^2} \int d^2 y {1 \over \sinh^2 y_1} 
\eea
where $u=0$ has been enforced and $\tilde{f}_6^2$ is the $AdS_6$ warp factor in string frame, to be evaluated at some yet-undetermined point $x$ on the boundary $\p\Sigma$. We work in Euclidean signature and take the $y_0$ coordinate to be such that $y_0 \sim y_0 + 2 \pi$. The remaining integral is divergent, but as usual we may regularize and perform holographic renormalization \cite{Skenderis:2002wp,Bianchi:2001kw,Balasubramanian:1999re}. In particular, noting that 
\bea
\int_\epsilon^\infty  {d y_1 \over \sinh^2 y_1} = {1 \over \epsilon} - 1 +{\epsilon \over 3 } + O(\epsilon^3)
\eea
we find renormalized action
\bea
\label{needforwl}
S' =  \tilde{f}_6^2 / \ell_s^2
\eea 
Since this can also be obtained via a Legendre transform of $y_1$, we will often refer to it as the energy. From now on we take $\ell_s^2 = 1$.

\subsection{Boundary Values of $\tilde{f}_6^2$}
We now study the shape of the energy profile of our embedded probe string. In particular, we would like to know the locations of the energy minima along the boundary of $\Sigma$. As noted before, we will only consider the case of $\Sigma$ being the upper-half plane plus a point at infinity, in which case $\p \Sigma$ is the real line parameterized by a real coordinate $x$.

While across the full surface $\Sigma$ the analysis of the function $\tilde{f}_6^2$ is rather intractable, on the boundary of $\p \Sigma$ things simplify significantly. In Appendix B, an exact boundary expression for $\tilde{f}_6^2$ is obtained - we reproduce the result here for convenience,
\bea
\label{sff62bound}
\tilde{f}_6^2 = e^\phi f_6^2  = 4 \sqrt{2} \, \sqrt{ \left(\mathrm{Re}\, \f_+\right)^2 + {3 \over 2} {\cG_{(1)} \over \k_{(1)}^2} \, \left(\mathrm{Re}\, \f_+' \right)^2}
\eea
The function $\f_+(x)$ is the leading order coefficient in the $(i y)^n$ power series expansion of $\cA_+(w)$, with $w= x + i y$. Likewise, the functions $\k_{(1)}^2$ and $\cG_{(1)}$ are the leading order coefficients in the power series expansions of $\k^2$ and $\cG$, respectively.  
The expression (\ref{sff62bound}) may now be explicitly extremized with respect to position $x$ on the boundary. In particular, one sees immediately that the derivative $\p_x \tilde{f}_6^2$ has an overall multiplicative factor of $\mathrm{Re}\, \f_+' $, and thus that
\bea 
\label{realcond}
\mathrm{Re}\, \f_+'  = 0
\eea
gives obvious extrema.\footnote{One might worry that the function $\k_{(1)}^2 $ also vanishes when $\mathrm{Re} \,\f_+' = 0$,  making our `extrema'  singular. However, this does not occur. As per (\ref{newreqs}), the function $\k_{(1)}^2$ is constrained to be positive definite by regularity conditions \cite{Gutperle:2017}.} In fact, these are not just any extrema - they are the global minima of the function $\tilde{f}_6^2$!

We illustrate this with some examples. To begin, consider a configuration of four semi-infinite external branes with symmetric charge assignments. The $(p,q)$-charges of the external branes, as well as the positions of their corresponding poles along the boundary, can be summarized by the following data, 
 \bea
 \label{easyembedding}
\hspace{0.5 in} p= {8 \over 3} N \, \left(-1, \, \,1,\,\,1,-1\right) \hspace{1.3 in} q = {8 \over 3} N \,\left(-1,-1,\,\,1,\,\,1\right)\hspace{0.5 in}
 \no\\\no\\ r = \left(1 - \sqrt{2},-1 + \sqrt{2},\,\,1 + \sqrt{2},-1 - \sqrt{2}\right)\hspace{1.3 in}
 \eea
 with $N$ an arbitrary integer, and the factors of $8/3$ included for convenience. 

\begin{figure}
\centering
\includegraphics[scale=0.6]{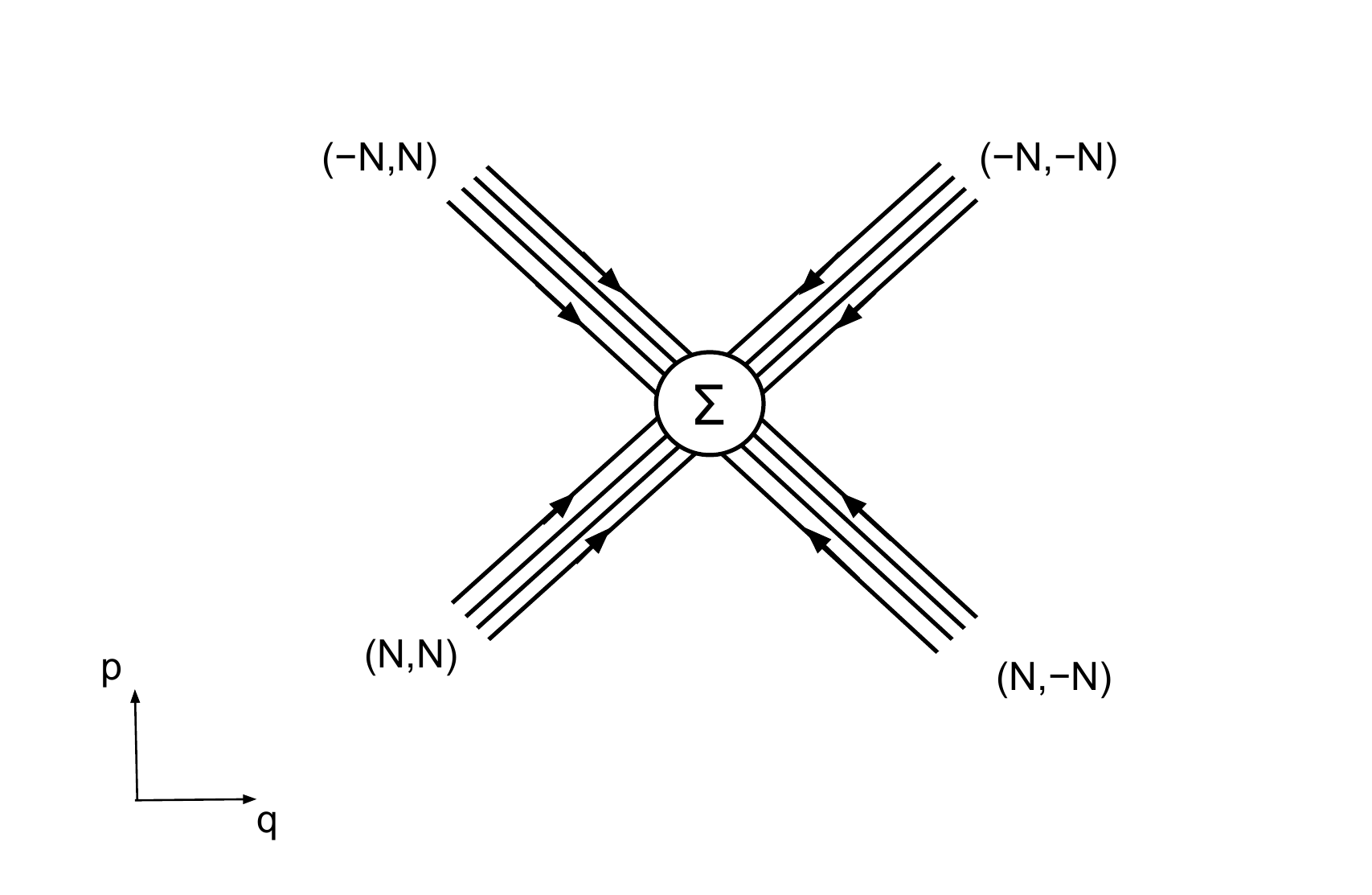}
\caption{The conformal limit of a brane web with four external $(p,q)$ five-branes and charge assignments of (\ref{easyembedding}). Factors of $8/3$ have been suppressed in labelling. The $(p,q)$ are taken to have orientation inwards towards the vertex. The circle at the singular intersection point of the web represents the region where the supergravity solutions are expected to hold. When $\Sigma$ is taken to be the unit disc, it can roughly be thought of as being in the plane of the web and located at this center point.}
\label{fig:1}
\end{figure}

The brane web picture corresponding to these charges is shown in Figure \ref{fig:1}. The entries in the $p$ and $q$ arrays of (\ref{easyembedding}) give the $(p,q)$-charges of each of the semi-infinite external branes, starting from the top right corner and proceeding clockwise around the web. These charges are always given with an inwards-flowing convention. Crucially, $(p,q)$-charge conservation is obeyed at the vertex. Note that in our conventions, D5-branes are (0,1)-branes and appear horizontal in the plane of the web, while NS5-branes are (1,0)-branes and are vertical in the plane. S-duality transformations correspond to rotations in the plane of the web. 

\begin{figure}
\centering
\includegraphics[scale=1]{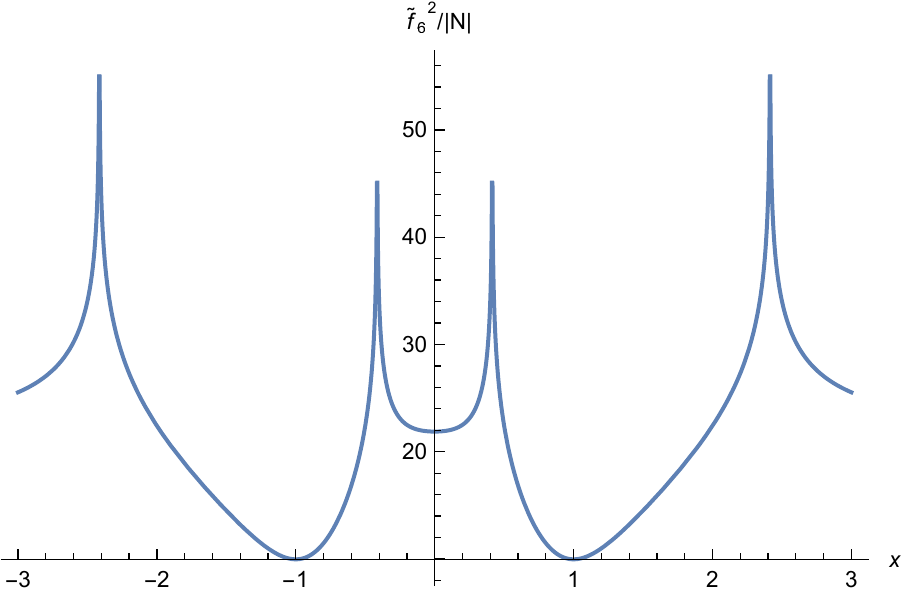}
\caption{The function $\tilde{f}_6^2/ |N|$, evaluated on the boundary of $\Sigma$ for the symmetric four external brane background of (\ref{easyembedding}). Note the global minima at $x=\pm1$, as well as the divergence at each of the poles.}
\label{fig:4sym}
\end{figure}

The  $r$ array in (\ref{easyembedding}) gives the $x$ location of the poles on $\p\Sigma$ corresponding to each of the branes, again beginning with the top right corner and proceeding clockwise. Three of the four poles were chosen freely using the $SL(2, \RR)$ symmetry of $\Sigma$. The fourth was then determined via the regularity conditions (\ref{regularity1}). For completeness, we also note that these regularity conditions can be used to obtain the complex constant $\cA^0$, giving $ \cA^0 = i \, N \, \log\left(3 + 2 \sqrt{2} \right)$ for this case.

The boundary value of the function $\tilde{f}_6^2/ |N|$ for this setup is plotted in Figure \ref{fig:4sym}.\footnote{Note that the function $\tilde{f}_6^2/ |N|$ is independent of $N$ due to the overall scaling results shown in (\ref{overallscale}).} From this plot, it is clear that the lowest energy configurations correspond to the minima at $x = \pm 1$, and we may confirm that this is consistent with the condition
 \bea
\label{symmemb}
\mathrm{Re} \, \f_+' = {4 \sqrt{2} N \over \prod_i (x - r_i)}  \,\left( x^2 - 1\right) = 0
\eea
whose solutions are also at $x = \pm 1 $.

Let us now analyze a more general arrangement of four external branes. In particular, we take 
 \bea
 \label{harderembedding}
 \hspace{0.5 in}p= {8 \over 3} \,\left(-p_1, \, \,p_2,\,\,p_1,-p_2\right) \hspace{1.3  in} q = {8 \over 3} \, \left (-q_1,-q_2 ,\,\,q_1,\,\,q_2\right)\hspace{0.5 in}
 \no\\\no\\ r = \left(1 - \sqrt{2},-1 + \sqrt{2},\,\,1 + \sqrt{2},-1 - \sqrt{2}\right)\hspace{1.3 in}
 \eea
This is still not the most general case with four external branes, but it will be general enough for us to observe many interesting properties. It also has the nice simplifying property that the locations of the four poles on $\p\Sigma$ can be taken exactly as before, regardless of the choice of the four free parameters $p_1$, $p_2$, $q_1$, and $q_2$. For completeness, we note that the regularity condition in this case gives
\bea
\cA^0 =\left(p_2 - p_1 \right) \log \left[1 + \sqrt{2} \right] + i \left(q_2 + q_1 \right)  \log  \left[1 + \sqrt{2} \right] 
\eea

\begin{figure}
\centering
\includegraphics[scale=1]{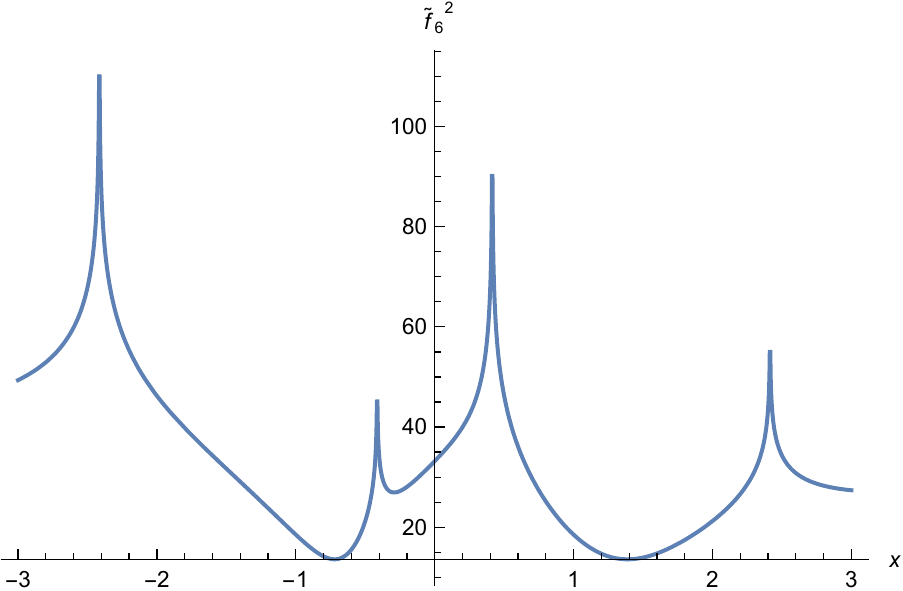}
\caption{The function $\tilde{f}_6^2$, evaluated on the boundary of $\Sigma$ for the semi-symmetric four external brane background of (\ref{harderembedding}) with $(p_1, p_2, q_1, q_2) = (1,2,1,3)$. Note the global minima at $ x_{\pm}={1 \over 3} \left(1 \pm \sqrt{10} \right)$.}
\label{fig:4anti}
\end{figure}

The global minima for this setup can again be obtained via the condition $\mathrm{Re}\,\f'_+=0$, which has two solutions, 
 \bea
 \label{4polemin}
x_{\pm} = {p_2 - p_1 \pm \sqrt{2\left(p_1^2 + p_2^2 \right)}\over p_1 + p_2}
\eea
For $p_1 = p_2 = q_1 = q_2 = N$, this reduces to the previous configuration. A less symmetric case with $(p_1 , p_2 , q_1 , q_2) = (1,2,1,3)$, is plotted in Figure \ref{fig:4anti}. The minimal energy embeddings in this particular case are at
\bea
x_{\pm} = {1 \over 3} \left(1 \pm \sqrt{10} \right)
\eea

\begin{figure}
\centering
\includegraphics[scale=1]{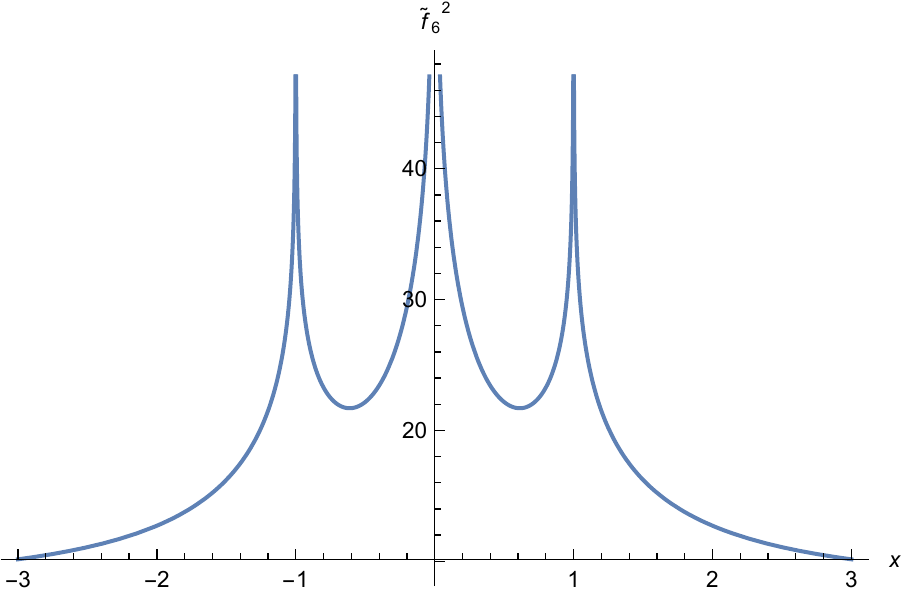}
\caption{The function $\tilde{f}_6^2$, evaluated on the boundary of $\Sigma$ for the three external brane background with $(p_1, p_2, q_1, q_2) = (1, 1, -1, 1)$. Note that the global minimum is at infinity. This is an example of what will be referred to as a Sector 1 probe string in Section 5.1.}
\label{fig:4}
\end{figure}

Finally, we consider a case with three external branes. The most general three-pole configuration is specified by the following data,
 \bea
 \label{3poledata}
 p = {8 \over 3} \, \left( p_1, -(p_1 + p_2), \,\,p_2\right) \hspace{0.5 in} q = {8 \over 3} \left( q_1, -(q_1+q_2), \,\,q_2\right) \hspace{0.5 in} r = \left(1,\,\,0,-1 \right)\hspace{0.3 in}
 \eea
where the location of the poles has been fixed by $SL(2, \RR)$ symmetry.  One finds via regularity that 
\bea
\cA^0 = -(p_1 + p_2) \log 2 + i (q_1+q_2) \log 2
\eea 
The global minimum of $\tilde{f}_6^2$ in such backgrounds is obtained by solving $\mathrm{Re}\, \f_+' = 0$, which has one solution 
\bea
\label{3polemin}
x = {p_1 + p_2 \over p_2 - p_1}
\eea
This is illustrated in Figure \ref{fig:4} for the case $p_1 = p_2 = q_2 = 1$ and  $q_1 = -1$. Note that in this case the global minimum is at infinity, in accordance with (\ref{3polemin}).

In Section 5, we will try to explain many of the aspects of these profiles via the brane web interpretation. Before doing so though, we will want to understand other aspects of these embeddings, such as whether or not they preserve supersymmetries. 
\subsection{Supersymmetry Conditions}
 
We now ask which embeddings preserve half of the background supersymmetries, as required for the 1/2-BPS probes that we are interested in. We may approach this problem in two ways. The first is by noting that two projection conditions $\e =  \Gamma_{(i)} \,\e$ for $i=1,2$ are compatible only if the matrices $\Gamma_{(1)}$ and $\Gamma_{(2)}$ commute. Actually in Type IIB, the projection conditions (coming for example from $\k$-symmetry) are of the form $\e =  \Gamma_{(i)}\, \e^*$, so we must take into account the complex conjugation. Thus we are interested in compatibility conditions of the form
\bea
\label{SUSYcompatibility}
\Gamma_{(1)} \Gamma_{(2)}^* - \Gamma_{(2)} \Gamma_{(1)}^* = 0
\eea
We now aim to construct the relevant matrices $\Gamma_{(i)}\,$. First we examine the supersymmetries preserved by the background geometry. As per \cite{DHoker:2016ujz}, we begin by decomposing the supersymmetry parameter $\epsilon$ as 
\bea
\label{edecomp}
\epsilon = \sum_{\eta_1, \eta_2 = \pm} \chi^{\eta_1 \eta_2} \otimes \zeta_{\eta_1\eta_2}
\eea
where the $ \chi$ are Killing spinors on $AdS_6 \times S^2$ and the $\zeta$ are remaining 2-component spinors. 
With this decomposition, the dilatino equation \cite{DHoker:2016ujz} reduces to a constraint on $\zeta$, 
\bea
\label{dilatino}
\zeta_{\eta_1 \eta_2} = \eta_2\left( \matrix{0 & \bar \alpha / \bar \beta \cr \beta / \alpha & 0 \cr}\right) \, \zeta^*_{\eta_1 \eta_2}
\eea
The matrix elements above are determined by the gravitino equations, giving 
\bea
\label{abratios}
\left({\beta \over \alpha}\right)^2 = {B \, \bar \kappa_+ + \bar \kappa_- \over \bar \kappa_+ + \bar B \,\bar \kappa_-}\hspace{1 in}\left({\bar \alpha \over \bar \beta}\right)^2 = {\kappa_+ + B\, \kappa_- \over \bar B\, \kappa_+ + \kappa_-}
\eea
where $\kappa_\pm \equiv \p_w \cA_\pm$ and B is the axion-dilaton of (\ref{axdil}).

We now examine $\left(\chi^{\eta_1\, \eta_2}\right)^*$. In fact, we may choose a real basis for the Killing spinors $\chi\,$ such that complex conjugation has no effect. This can be seen by recalling that the Killing spinors $\chi\,$ satisfy 
 \bea
\left(\chi^{\eta_1\, \eta_2}\right)^* = \eta_2 \left(B_{(1)} \otimes B_{(2)} \right) \chi^{\eta_1\, \eta_2}
 \eea
 with $B_{(1)} \otimes B_{(2)}$ a real matrix (see (\ref{compconmat}) in Appendix A). Adding/subtracting the above equation with its complex conjugate then gives 
 \bea
 \mathrm{Re}\,\chi^{\eta_1 \,\eta_2} = \eta_2 \left(B_{(1)} \otimes B_{(2)} \right)  \mathrm{Re}\,\chi^{\eta_1\, \eta_2} \hspace{0.6 in}  \mathrm{Im}\,\chi^{\eta_1\, \eta_2} = - \eta_2 \left(B_{(1)} \otimes B_{(2)} \right)  \mathrm{Im}\,\chi^{\eta_1\, \eta_2} 
 \eea
Thus $\{\mathrm{Re}\,\chi \, , i \, \mathrm{Im}\,\chi\}$ for $\eta_2 = 1$ and $\{i \,\mathrm{Re}\,\chi \, , \, \mathrm{Im}\,\chi\}$ for $\eta_2 = -1$ constitute a real basis, which we work in from now on. 
 Using this and (\ref{dilatino}), we find that the background geometry preserves supersymmetries satisfying $\epsilon = \Gamma_{(1)} \epsilon^* $ with
 \bea
\Gamma_{(1)} = \mathds{1}_8 \otimes \mathds{1}_2\otimes \left( \matrix{0 &\bar \alpha /  \bar \beta \cr \beta /  \alpha & 0 \cr}\right) 
 \eea
As expected, this condition is traceless and (taking the complex conjugation into account)  squares to one, and so it preserves half of the supersymmetries. 

We may now examine the supersymmetries which are preserved by the probe F1-string embedding. The relevant condition follows from familiar $\k$-symmetry arguments \cite{Green:1983wt,Grisaru:1985fv}, and is just $\epsilon = \Gamma^{01} \epsilon^* $. Hence our second matrix is
\bea
\label{fundstringproj}
\Gamma_{(2)} = \Gamma^{01}= -\left(\sigma^3 \otimes \mathds{1}_2 \otimes \mathds{1}_2 \right)\otimes \mathds{1}_2\otimes \mathds{1}_2 
\eea
which is again traceless and of square one. Requiring that it be compatible with the background supersymmetry condition then amounts to enforcing (\ref{SUSYcompatibility}). This leads to the simple requirement that $\mathrm{Im} \, {\a \over \b} =0$, i.e. that 
\bea
\label{11111}
{\a \bar \b \over \bar \a \b} = 1
\eea
For the fundamental string embeddings we have been exploring here, the probe strings are restricted to the boundary of $\Sigma$, where we have $\a \bar \a = \b \bar \b$.\footnote{The definitions (\ref{abratios}) and the fact that $\k_\pm = - \bar \k_\mp$ on the boundary give us that $\a \bar \a = \pm \b \bar \b$. We may check that the plus sign is the correct choice by comparing to equation (3.46) in \cite{DHoker:2016ujz}, i.e. $f_2 \sim (\a \bar \a - \b \bar \b)$ and  $f_6 \sim (\a \bar \a + \b \bar \b)$, and recalling that $f_2$ vanishes identically on the boundary while $f_6$ does not.} Hence it suffices to impose the simpler condition 
\bea
\left({\a \over \b}\right)^2 = 1 
\eea

To see what restrictions this puts on the allowed embeddings, we use (\ref{abratios}) and the boundary expansion for $B$ in (\ref{Bbound}) to obtain an explicit boundary expression for $({\a / \b})^2$. The result is 
\bea
\left({\a \over \b}\right)^2 = {- i \,\bar \f_+' \, \sqrt{6 \cG_{(1)}} +\, 2\, \bar \f_+ \k_{(1)} \over  i \,\f_+' \, \sqrt{6 \cG_{(1)}} + \,2\, \f_+ \k_{(1)}}
\eea
Setting this equal to $1$ then amounts to requiring that 
\bea
\label{finalSUSYcond}
\mathrm{Im} \left[i \f_+' \, \sqrt{6 \cG_{(1)}} + 2 \,\f_+ \k_{(1)} \right] = 0  \hspace{0.4 in} \Rightarrow\hspace{0.4 in}\mathrm{Im} \, \f_+ =  - \sqrt{3\, \cG_{(1)} \over 2 \,\k_{(1)}^2} \,\mathrm{Re} \, \f_+'
\eea
This gives us an explicit condition on which embedding locations $x$ of our string preserve half of the supersymmetries. This form of the supersymmetry condition will be used often in what follows.

The second way to derive this same result is by first solving the Killing spinor equations for the form of the background Killing spinor \cite{Lu:1996rhb,Lu:1998nu}. This is achieved via a fairly standard calculation, which in this case gives the result 
\bea
\e = (1 + i \nu) \left[e^{\half \g_1 y_1} e^{\half \g_0 y_0} \otimes \mathds{1}_2 \otimes\mathds{1}_2 -  e^{-\half \g_1 y_1} e^{-\half \g_0 y_0} \otimes \mathds{1}_2 \otimes \sigma^3\right]  \e_1 \otimes \e_2 \otimes \left(\matrix{\bar \a \cr \b} \right)\no
\eea
Here $\e_1$ and $\e_2$ are constant spinors of dimension 8 and 2, respectively, and $\nu= \pm 1$. One may now impose the condition (\ref{fundstringproj}) on this Killing spinor. This imposes constraints on the constant spinor $\e_1$, but also on the final, background-dependent term. Requiring half of the supersymmetries to be preserved then results in the condition (\ref{11111}) found before.

For embeddings preserving supersymmetry, the boundary expression (\ref{sff62bound}) for the string-frame metric factor $\tilde{f}_6^2$ simplifies beautifully to 
\bea
\label{simplesff62}
\tilde{f}_6^2 = 4 \sqrt{2} \, \sqrt{ \left(\mathrm{Re}\, \f_+\right)^2 +\left( \mathrm{Im} \, \f_+ \right)^2} =4 \sqrt{2} \, \left|\f_+\right|
\eea
It should be kept in mind that this expression holds only at the points $x$ at which (\ref{finalSUSYcond}) is satisfied - it should not be thought of as a new functional form for $\tilde{f}_6^2$ at generic points.

\subsection{Minimal Energy vs. Supersymmetric Embeddings}

An immediate consequence of the supersymmetry condition (\ref{finalSUSYcond}) is the fact that the globally minimal energy F1-string embeddings are not necessarily supersymmetric. Indeed, in Section 3.1 the globally minimal energy embeddings were found to be at $x$ satisfying $\mathrm{Re} \,\f_+' = 0$. But this condition is generally not compatible with (\ref{finalSUSYcond}). We see that what is required for minimal energy embeddings to be supersymmetric is that $\mathrm{Im} \, \f_+ $ also vanish at the same $x$.  Below, we will use this condition to explore the compatibility of the minimal energy and supersymmetric embedding conditions for the four- and three-external brane cases introduced in Section 3.1.

\begin{figure}
\centering
\begin{minipage}{.5 \textwidth} 
\centering
\includegraphics[scale=0.55]{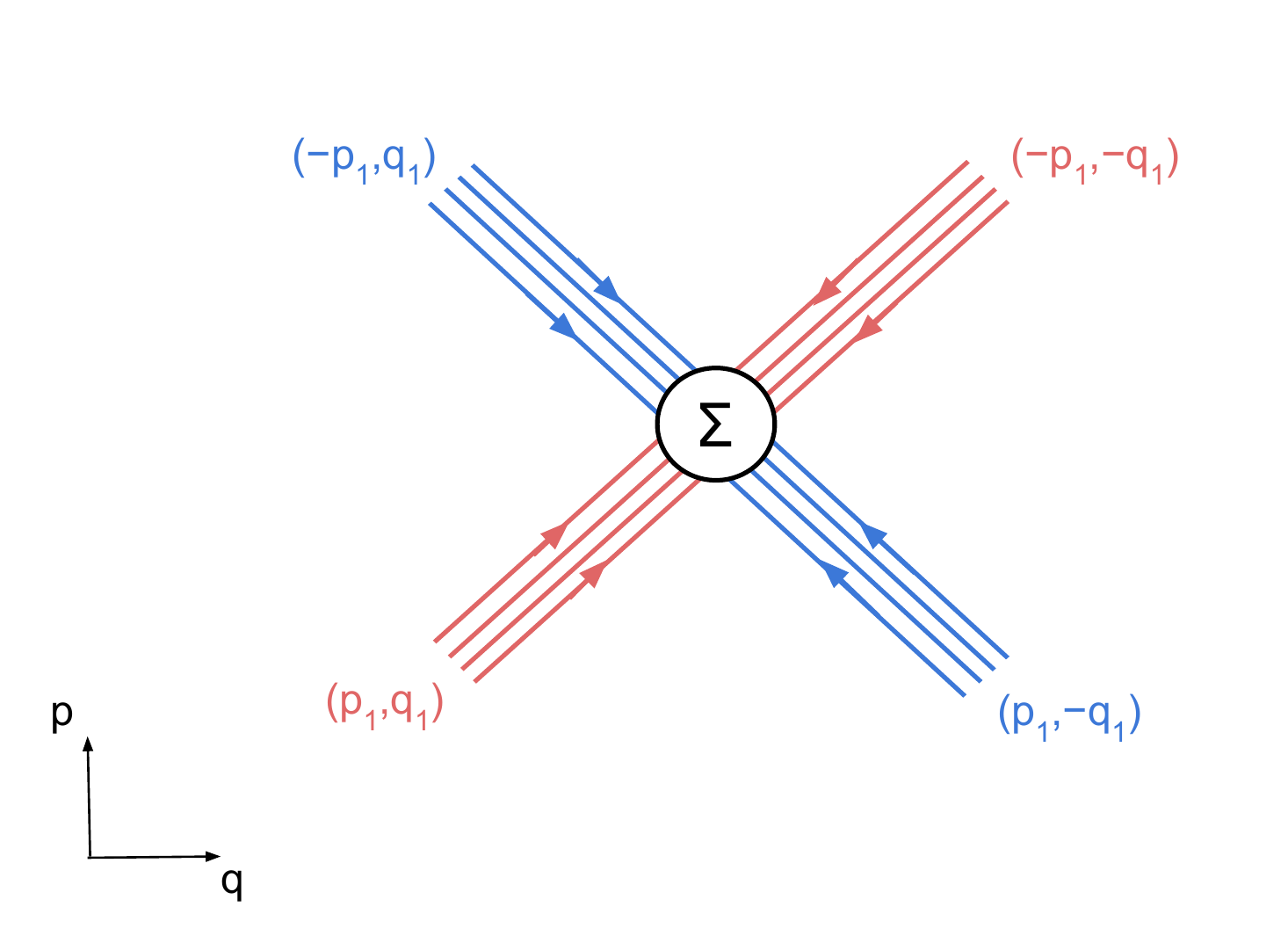}
\end{minipage}%
\begin{minipage}{.5 \textwidth} 
\centering
\includegraphics[scale=0.55]{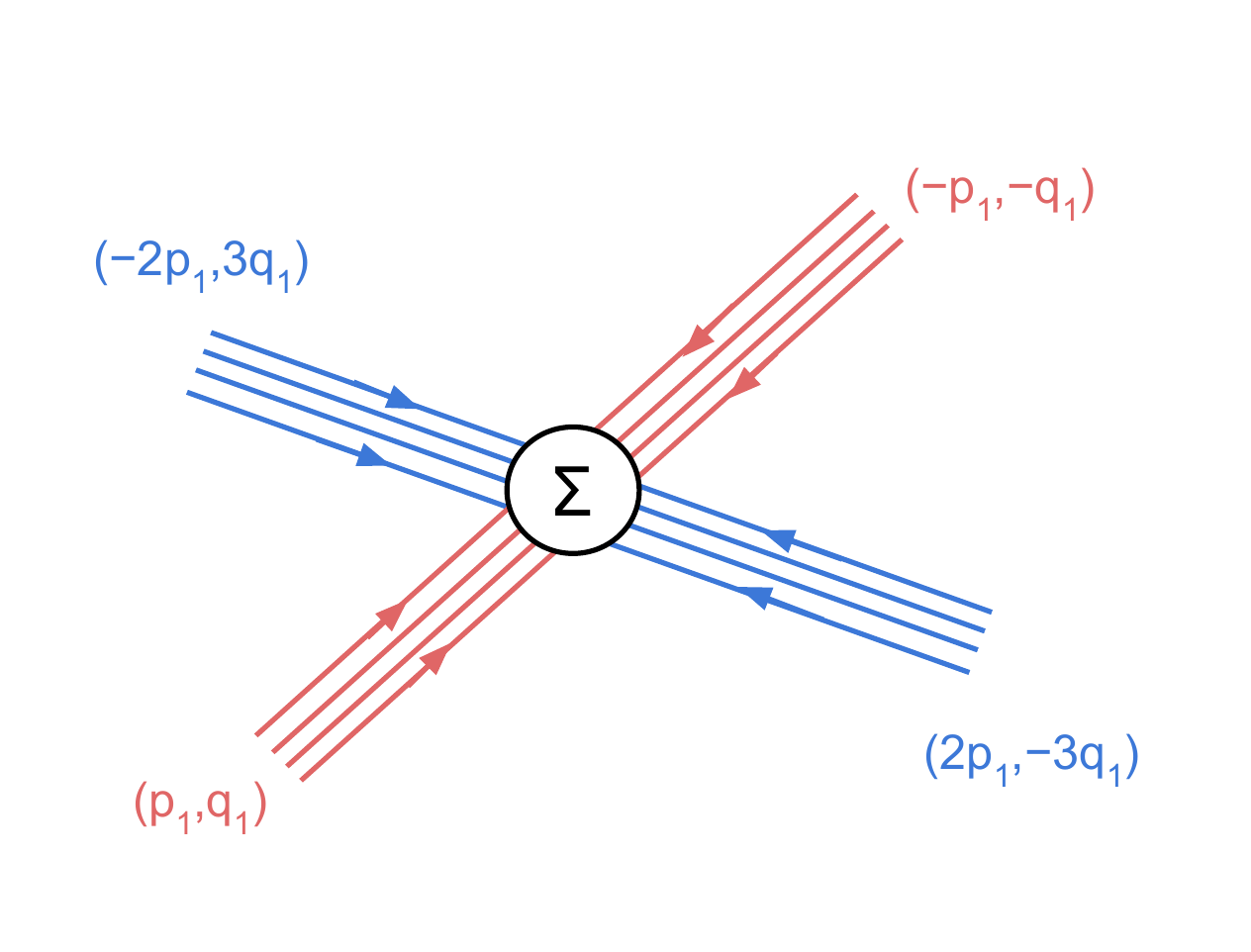}
\end{minipage}%
\caption{A schematic picture of two known classes of four-external brane backgrounds allowing supersymmetric, minimal energy F1-string embeddings; (a) has  $p_2 =  p_1$ and $q_2 = q_1$, while (b) has $p_2 =  2 \, p_1$ and $q_2 = 3\, q_1$. Charges  $p_2 =  -2 \, p_1$ and $q_2 = 3\, q_1$ are also allowed, though not pictured.}
\label{allowedwebs}
\end{figure}

We begin with the four-external brane configuration (\ref{harderembedding}), which contains the more symmetric configuration (\ref{easyembedding}) as a special case. To see which backgrounds admit supersymmetric, minimal energy embeddings, we insert the extrema $x_\pm$ of  (\ref{4polemin}) into $\mathrm{Im} \, \f_+ $ and require that it vanish. This gives the following constraint on the $(p_i,q_i)$-charges of the external branes, 
\bea
\label{4poleconst}
\left|p_1 \mp \sqrt{p_1^2 + p_2^2}\,\right|^{q_2} \,\left|p_2 \pm \sqrt{p_1^2 + p_2^2}\,\right|^{q_1} = \left|p_2\right|^{q_2} \left|p_1\right|^{q_1}
\eea
Without loss of generality, we can say that the two $p_i$ charges are related by $p_2 = n p_1$ for $n \in \mathbb{Q}$. Then the above is solved by $q_2 = {\mathrm{ArcSinh}\,n \over \mathrm{ArcCsch}\, n }\,q_1$. Because the ratio of any two brane charges should be a rational number, we conclude that in these four-pole cases a supersymmetric minimal energy embedding is obtainable only if 
\bea
\label{rationalcondition}
  {\mathrm{ArcSinh}\,n \over \mathrm{ArcCsch}\, n }\in \QQ
\eea
The simplest non-degenerate $n$ satisfying this condition are $n= 1$, $\pm2$, and $\pm\half$. Since no other solution to (\ref{rationalcondition}) has yet been identified, these may actually be \textit{all} the possiblities, though we do not make any strong claim of this. In any case, for $n=1$ we have that $p_2 = p_1$ and $q_2 = q_1$.  For $n=\pm 2$ we find that $p_2 = \pm 2 \, p_1 $ and $ q_2 = 3\, q_1$. Finally for $n=\pm\half$ we have $p_2 = \pm{\half}  p_1$ and $ q_2 = {1 \over3} q_1$, which we recognize to be the $n=2$ case with the charges on each side interchanged. Thus we have identified two distinct classes of backgrounds allowing supersymmetric, minimal energy F1-string embeddings, depicted schematically in Figure \ref{allowedwebs}.

We now examine the case of a general configuration of three external branes, specified by  (\ref{3poledata}). We follow the same routine of plugging (\ref{3polemin}) into $\mathrm{Im} \, \f_+ $ and demanding that it vanish. This imposes the following condition on the charges, 
\bea
|p_1 + p_2|^{q_1 + q_2} = |p_1|^{q_1}|p_2|^{q_2}
\eea
Taking $p_2 = n p_1$ for $n \in \mathbb{Q}$ again, we find that $q_2 = {\log(n+1) \over \log(n) - \log(n+1)} \,q_1$. The only obvious solution giving a rational result for $q_2/q_1$ has $n=1$, in which case $p_2 =  p_1$ and $q_2 = - q_1$. 

\section{D1-String Embeddings}
\setcounter{equation}{0}
We now turn to the embedding of probe D1-strings, in the same manner as for the fundamental strings before. Since the D1-string is related to the fundamental string by S-duality, we expect its embeddings to be described in an analogous manner to those above. Thus we sketch only a broad outline of the relevant results here. 

\begin{figure}
\centering
\includegraphics[scale=1]{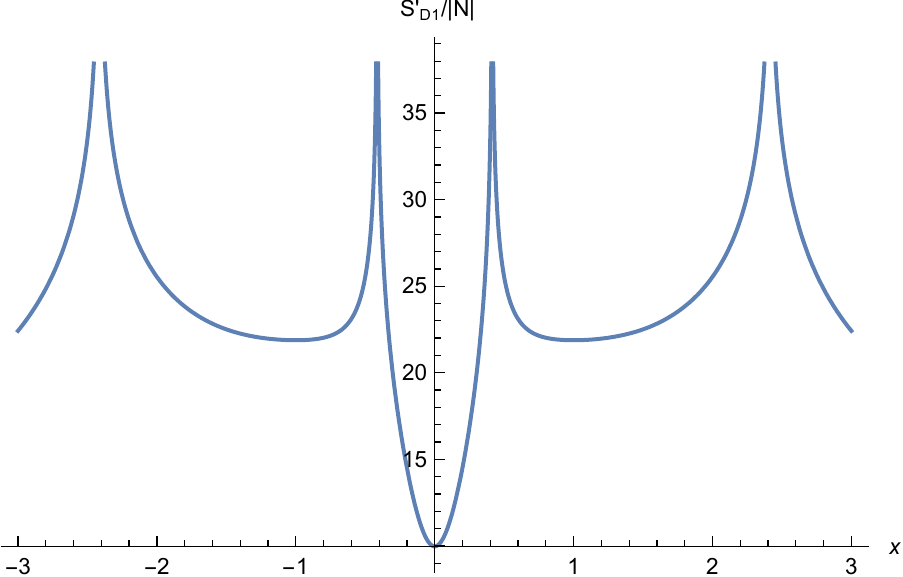}
\caption{The function $S'_{D1}/|N|$, evaluated on the boundary of $\Sigma$ for the symmetric four external brane background of (\ref{easyembedding}). Note that $S'_{D1}/|N|$ is independent of $N$ by the scaling relations of (\ref{overallscale}). }
\label{DString}
\end{figure}

To begin, the renormalized action for a D1-string wrapping $AdS_2$ takes the form 
\bea
\label{D1action}
S'_{D1} = \tilde{f}_6^2 \,\sqrt{e^{-4 \phi} + \chi^2 } 
\eea
An example of the resulting energy profile is shown in Figure \ref{DString}, which is plotted for the symmetric four-pole case of (\ref{easyembedding}). The global minima are now found to be at solutions of $\mathrm{Im} \, \f_+' = 0$. Note that this is the same condition as (\ref{realcond}) for the fundamental string, but with the real part replaced with imaginary. This can be easily explained via S-duality by noting that interchange of  $\mathrm{Re} \, \f_+' $ with $\mathrm{Im} \, \f_+' $ is just an interchange of  $\mathrm{Re} \, Z_+^k $ with $\mathrm{Im} \, Z_+^k $. These latter two quantities are related to the $(p,q)$-charges of the external branes via (\ref{pqcharges}), so this amounts to an exchange of $p$ and $q$ charges - exactly what we would expect in S-dualizing from an F1-string to a D1-string. 

The $\k$-symmetry condition on supersymmetries preserved by the probe is now $\e = \Gamma_{(2)} \e^*$ with 
\bea
\label{Dstringproj}
\Gamma_{(2)} = i  \Gamma^{01} = - i \left(\sigma^3 \otimes \mathds{1}_2 \otimes \mathds{1}_2 \right)\otimes \mathds{1}_2\otimes \mathds{1}_2
\eea
which is the usual result for D1-strings \cite{Aganagic:1996nn,Aganagic:1996pe}. Since this only differs from (\ref{fundstringproj}) by a factor of $i$, using either of the two methods described in Section 3.2 immediately implies that the condition on having a 1/2-BPS embedding is $\Re \, {\a \over \b} = 0$. More explicitly, the condition for a supersymmetric D1-string embedding is
\bea
\mathrm{Re} \, \f_+ =  \sqrt{3\, \cG_{(1)} \over 2 \,\k_{(1)}^2} \,\mathrm{Im} \, \f_+'
\eea
This can be compared with (\ref{finalSUSYcond}) and again understood in terms of S-duality. On embeddings satisfying this condition, the regularized action simplifies to 
\bea
S'_{D1} = 4 \sqrt{2} \, |\f_+|
\eea
exactly as in the case of the fundamental string. 

By the above, the condition that an embedding be both globally minimal energy and supersymmetric is that $\mathrm{Im} \, \f_+' = 0$, as well as $\mathrm{Re} \, \f_+= 0$. We can now identify the backgrounds which admit such embeddings in the same way as for the fundamental string. For example for backgrounds with four external branes of the form (\ref{harderembedding}), the minimal energy embeddings are at 
\bea
x_\pm = {q_1 + q_1 \pm \sqrt{2\left(q_1^2 + q_2^2 \right)}\over q_2 - q_1}
\eea
Plugging this into $\mathrm{Re} \, \f_+= 0$ and demanding that it vanish gives
\bea
\left|q_1 \mp \sqrt{q_1^2 + q_2^2}\,\right|^{p_2} \,\left|q_2 \pm \sqrt{q_1^2 + q_2^2}\,\right|^{p_1} = \left|q_2\right|^{p_2} \left|q_1\right|^{p_1}
\eea
This is the same constraint equation as (\ref{4poleconst}), but with the charges $p_i$ and $q_i$ interchanged. Thus all of the conclusions made earlier about acceptable and unacceptable backgrounds have direct analogs here. Similar comments hold for the three-pole backgrounds as well.

\section{(p,q)-Strings and Five-brane Webs}
\setcounter{equation}{0}
We now explore the connection between the $AdS_6 \times S^2 \times \Sigma$ supergravity solutions and $(p,q)$ five-brane webs. In particular, we will be able to explain qualitatively (and in some cases, quantitatively) many of the characteristic features of the energy profiles in Figures \ref{fig:4sym}, \ref{fig:4anti},  \ref{fig:4}, and \ref{DString} via microscopic considerations. In the first subsection below, we generalize from the F1 and D1 probe string embeddings studied thus far to the more general case of $(p,q)$ probe string embeddings. We will understand these embeddings in terms of the brane web picture, where the interactions of the $(p,q)$ probe string with the web occur via the formation of ``string junctions" with open strings ending on the branes of the web. The discussion in this subsection is largely qualitative, but conceptually quite useful. In the second subsection, we turn towards a more quantitative study of the correspondence between the supergravity solutions and the brane web picture. To do so, we will restrict ourselves to the original F1- and D1-string cases and try to reproduce some of the results of the previous two sections using brane web considerations. In particular, we will conjecture an equivalence between the total junction tension on the brane web side and a certain simple quantity on the supergravity side, and use this equivalence to recast the minimum energy and supersymmetry conditions found earlier as simple tension minimization conditions. 

\subsection{Probe $(p,q)$-Strings and String Junctions}

We now study embeddings of probe $(p,q)$-strings preserving the same maximal set of bosonic symmetries as the F1- and D1-strings before. On the supergravity side, the analysis would proceed as usual by finding the renormalized action for a probe $(p,q)$-string,
\bea
\label{pqstringaction}
S_{(p,q)}' = \tilde{f}_6^2 \sqrt{q^2 e^{- 4 \phi} + (p - q \chi)^2}
\eea
and following the same extremization procedure as outlined for the probe F1- and D1-strings. Instead of repeating this though, we will instead try to analyze things directly from the brane web picture. We begin by focusing on backgrounds with four external branes and charges given by (\ref{harderembedding}) - once the ideas have been introduced, we can move on to the slightly more involved three-pole case.

\begin{figure}
\centering
\begin{minipage}{.5 \textwidth} 
\centering
\includegraphics[scale=0.55]{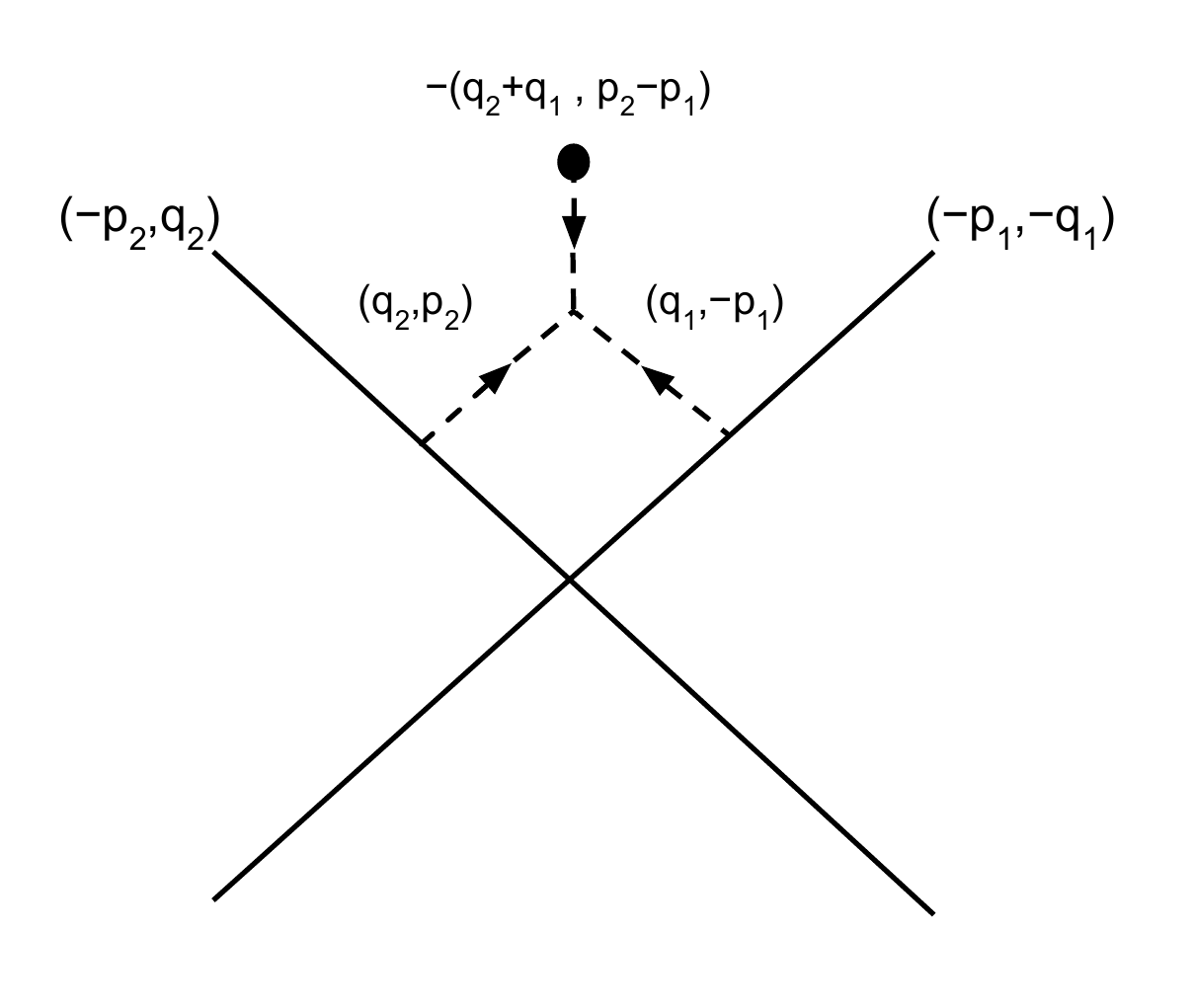}
\end{minipage}%
\begin{minipage}{.5 \textwidth} 
\centering
\includegraphics[scale=0.55]{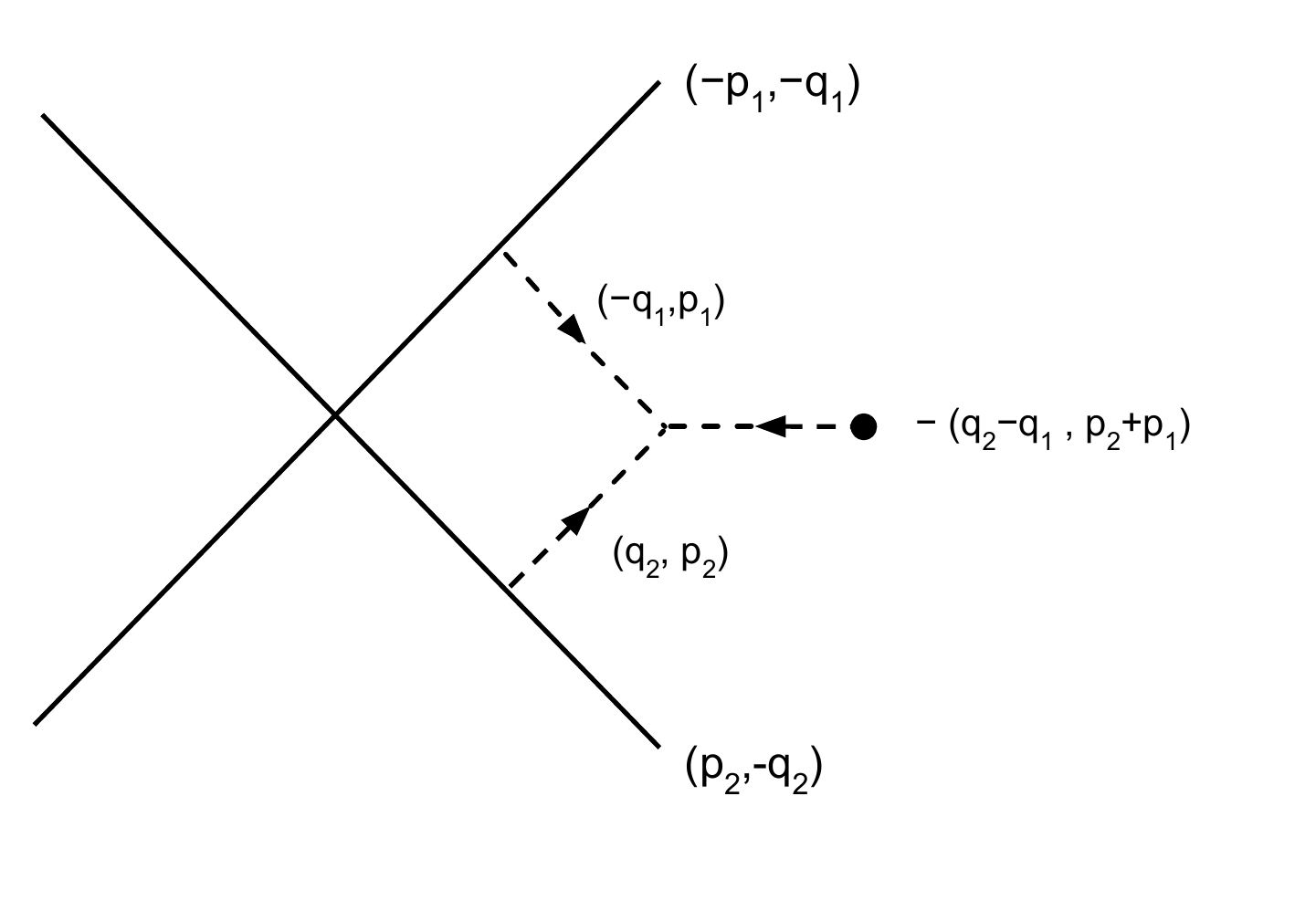}
\end{minipage}%
\caption{Junctions formed between three open strings (dashed lines). Two of the open strings are anchored on the brane web, and the third is anchored on the probe string (black dot). Each ($p_i, q_i$) brane has a $\pm$($-q_i,p_i$)-string living on it. Two such strings can meet and form a junction with the probe string if the charge of the probe string is consistent with charge conservation at the vertex. Just as for the brane web itself, we choose an inward-flowing charge convention.}
\label{fig:branejunc}
\end{figure}

Start by recalling that the probe strings are taken to be single points on the boundary of $\Sigma$. As will be seen more clearly later, the Riemann surface $\Sigma$ can be thought of as being ``located" at the singular intersection point of the external branes of the web (see Figure \ref{fig:1}), so we can also interpret the probe strings as single points embedded in the plane of the web. Though these embedding points are all localized at the intersection point of the web, we will often draw them as being located some finite distance away from it - see for example Figure \ref{fig:branejunc}. One important exception to the statement that all probe strings are located at the intersection point is the case in which a probe string is moved very close to a pole on the boundary of $\Sigma$. In this case, the probe string should be thought of as leaving the intersection point and moving off to infinity along one of the external branes of the web. This fact will be discussed more later.  

When probe $(p,q)$-strings are embedded in the web, they can interact with branes of the web via interactions with open strings ending on the branes. Note that only $\pm(-n, m)$-strings can be anchored on $(m,n)$-branes,  with the choice of overall sign amounting to the choice of orientation of the string, i.e. running to or from the brane. The relative negative sign in the charges of the string is necessary to make the string perpendicular to the brane.\footnote{If a line has slope $m$, then the line perpendicular to it has slope $- 1/m$.} Physically, this is the condition that there is no force longitudinal to the brane \cite{Aharony:1997bh, Aharony:1997ju}.


 Unless $(p,q)$ is an integer multiple of $\pm(-n,m)$, the probe $(p,q)$-string does not interact directly with such open strings. For the time being, we restrict ourselves to the case in which $(p,q)$ is not an integer multiple of $\pm(-n,m)$ for any $(m,n)$-brane in the web - we will return to this point momentarily. Then as just mentioned, the probe $(p,q)$-string cannot interact with any \textit{single} open string of the web. However, in certain cases there can be interactions with more than one open string. The mechanism allowing for such interactions are so-called ``string junctions", which have been well-studied in previous literature - see for example \cite{Schwarz:1996bh,Sen:1997xi, Dasgupta:1997pu}.\footnote{There is an important difference between the junctions discussed in the present paper and those discussed previously in the context of five-brane webs in \cite{Aharony:1997bh, Kol:1998cf}. The junctions in those papers were taken to be composed of open strings which lived within the face of the web - i.e. the open strings involved were anchored on internal branes. These junctions make no appearance in the present case since the embeddings being studied here are blind to any internal structure of the web - they probe only the exterior of the web, and thus the only junctions formed are with open strings anchored on external branes. In part, this may be because we have restricted to embeddings located on $\p\Sigma$. For embeddings in the interior of $\Sigma$, some data about the internal branes of the web may in fact be accessible - see footnote 10.} Such junctions may form whenever three strings meet at a vertex in a combination consistent with $(p,q)$-charge conservation. Two such junctions are shown in Figure \ref{fig:branejunc}, from which we see that for backgrounds given by charge assignments (\ref{harderembedding}), junctions can form whenever the charge of the probe string is an integer multiple of ${(q_2 \pm q_1, p_2 \mp p_1)}$. These are the only distinct three-string junctions that are available for such backgrounds. 
 
  Half of the brane web supersymmetries are preserved by this string configuration so long as every $(p,q)$-string has slope $s_{(p,q)}$ satisfying
 \bea
 \label{slopecondition}
 s_{(p,q)} = {\mathrm{Im}\left[{q + p \, \tau}\right] \over \mathrm{Re}\left[{q + p \, \tau}\right]}
 \eea 
 with $\tau$ the axion-dilaton \cite{Dasgupta:1997pu}. This is the same condition required of the $(p,q)$ 5-branes in the web. As is customary, when the web is drawn, $\tau$ is taken to be $i$ so that the slope simplifies to $s={p/q}$.

Note that regardless of the precise values of the background charges $(p_i ,q_i)$, so long as the number and location of semi-infinite external branes is kept the same the probe strings of charge $(q_2 \pm q_1, p_2 \mp p_1)$ will always have the same junction interactions with the web. Hence we should expect that all such probe strings possess the same energy profile (i.e. the energies are functionally equivalent up to overall constant factors). Indeed, this has been verified numerically for a large number of backgrounds.

\begin{figure}
\centering
\begin{minipage}{.5 \textwidth} 
\centering
\includegraphics[scale=0.55]{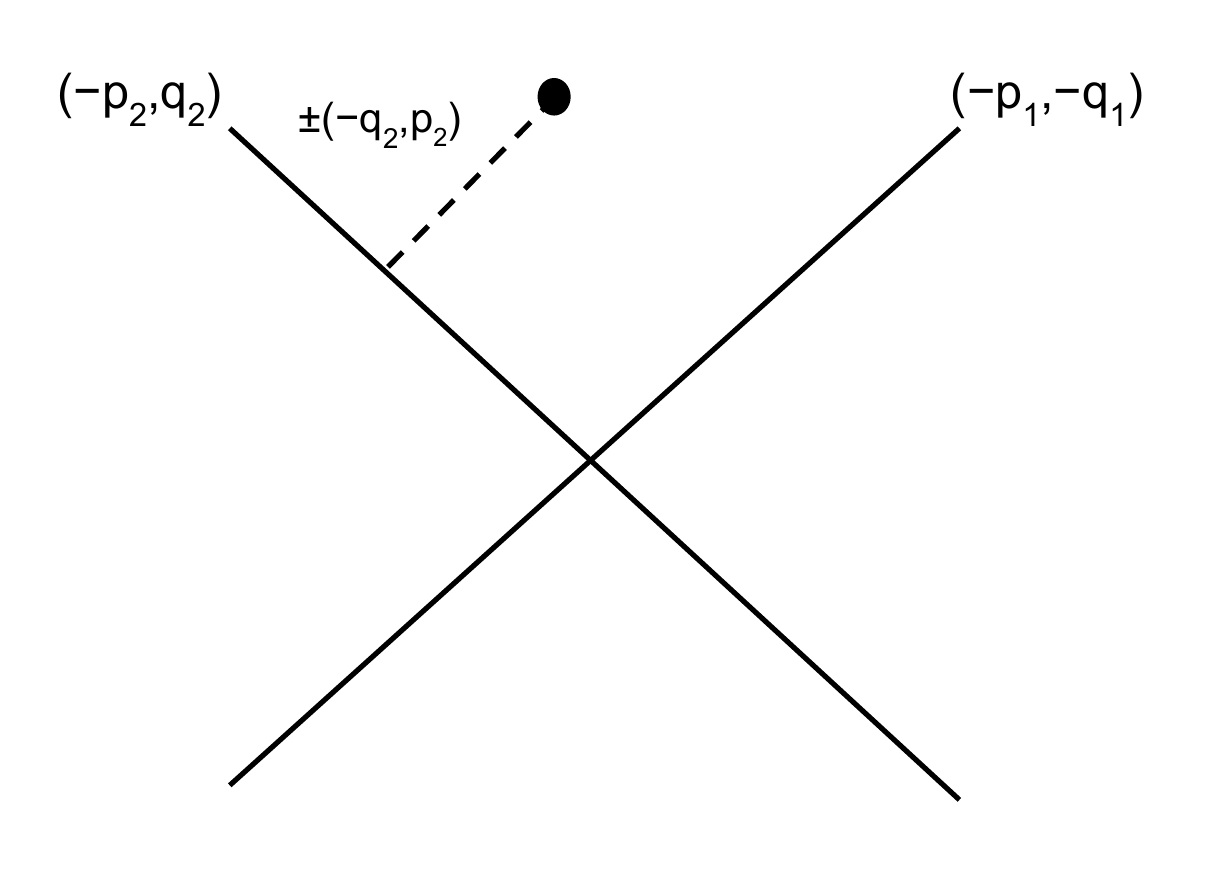}
\end{minipage}%
\begin{minipage}{.5 \textwidth} 
\centering
\includegraphics[scale=0.55]{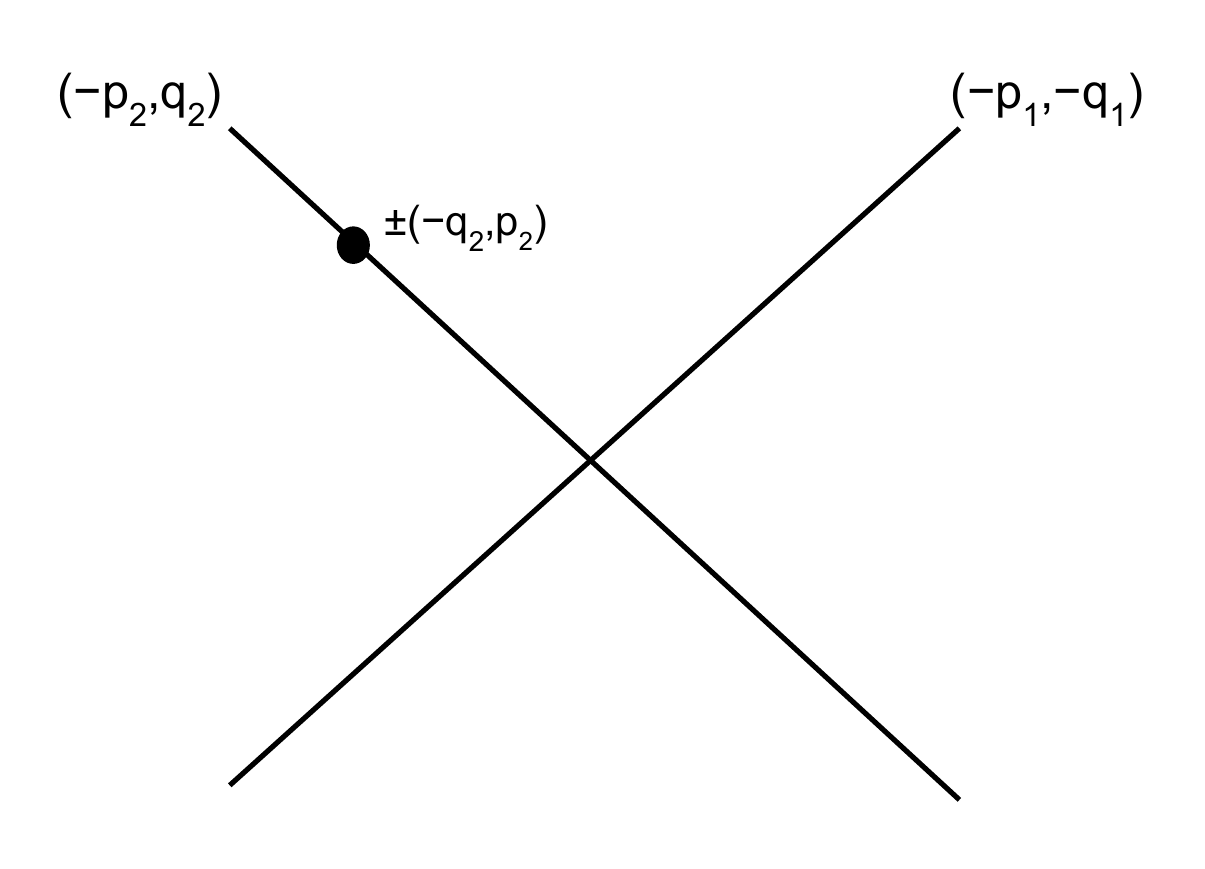}
\end{minipage}%
\caption{A probe $(-q_2,p_2)$-string in the presence of a $(p_2,q_2)$-brane. The probe string minimizes its energy when it's directly atop the brane, since in this case the mass of the open string stretching to it vanishes.}
\label{simplejunctions}
\end{figure}

\begin{figure}
\centering
\begin{minipage}{.5 \textwidth} 
\centering
\includegraphics[scale=0.82]{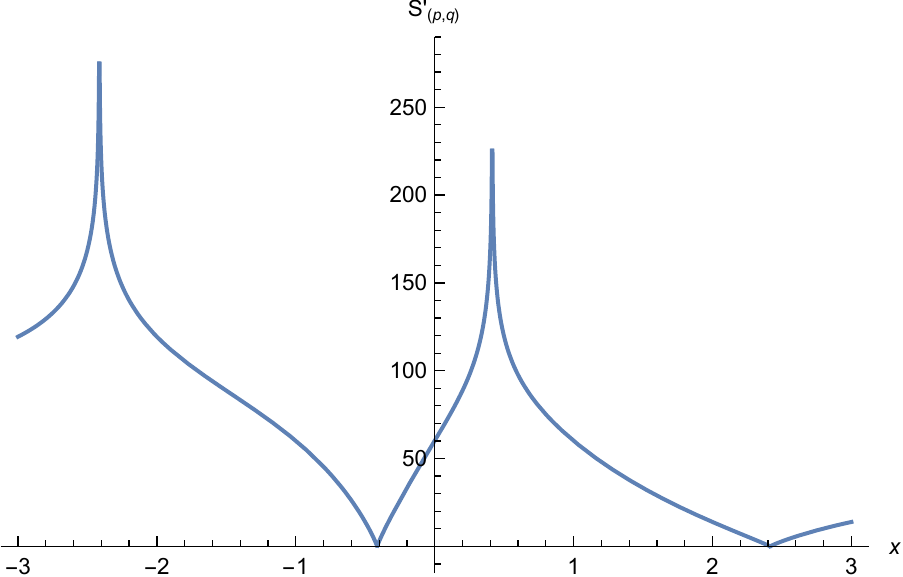}
\end{minipage}%
\begin{minipage}{.5 \textwidth} 
\centering
\includegraphics[scale=0.82]{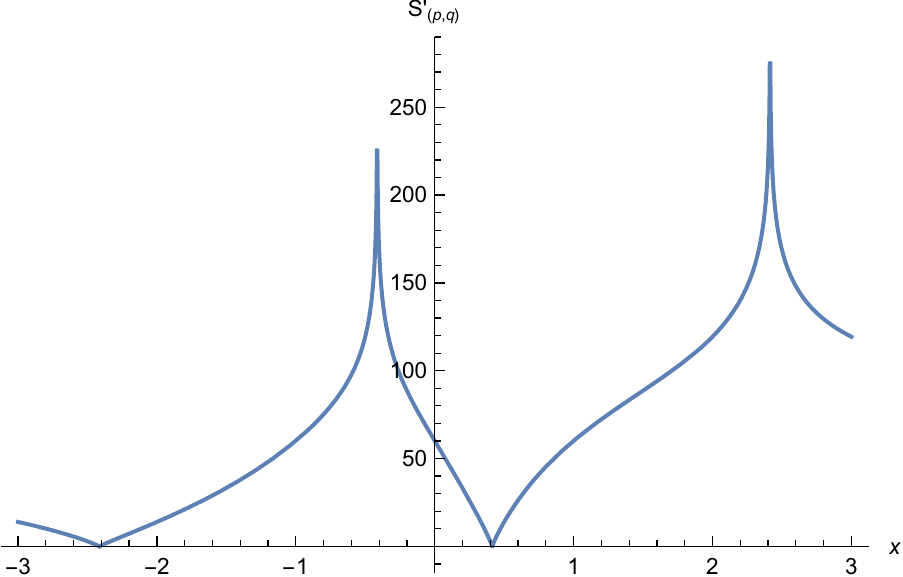}
\end{minipage}%
\caption{The energy profile of a probe string placed in a background of type (\ref{harderembedding}). Recall that branes of type $\pm (p_1, q_1)$ are located at $x = 1\pm \sqrt{2} $ and branes of type $\pm (p_2, -q_2)$ are located at $x = -1\pm \sqrt{2}$.  On the left, we have placed a ($-q_1,p_1$)-string in this background, and we see that its energy is minimized when the string is coincident with the $\pm (p_1, q_1)$ branes. On the right, we have placed a  ($q_2,p_2$)-string in the same background, and we see that its energy is now minimized when the string is coincident with the $\pm (p_2,- q_2)$ branes. For explicit results, we have taken $(p_1, p_2, q_1, q_2) = (1, 2, 1, 3)$.}
\label{simpleresults}
\end{figure}

We now turn to the case in which a probe $(-q_i,p_i)$-string is embedded in a web containing a $(p_i, q_i)$-brane. In this case, the probe string can interact directly with the open strings anchored on just the single brane.  On the supergravity side, one finds that in this case, the probe string action (\ref{pqstringaction}) is minimized when the probe string is located at the brane's corresponding pole on the boundary of $\Sigma$ - in fact, the energy there exactly vanishes. The brane web picture of this is clear: the probe $(-q_i,p_i)$-string locates itself directly atop the $(p_i,q_i)$-brane in $\Sigma$ so as to minimize the energy of the open string anchored on the brane and stretching to it. This situation is shown in Figure \ref{simplejunctions}, with energy profiles shown in Figure \ref{simpleresults}.

 \begin{figure}
\centering
\includegraphics[scale=0.55]{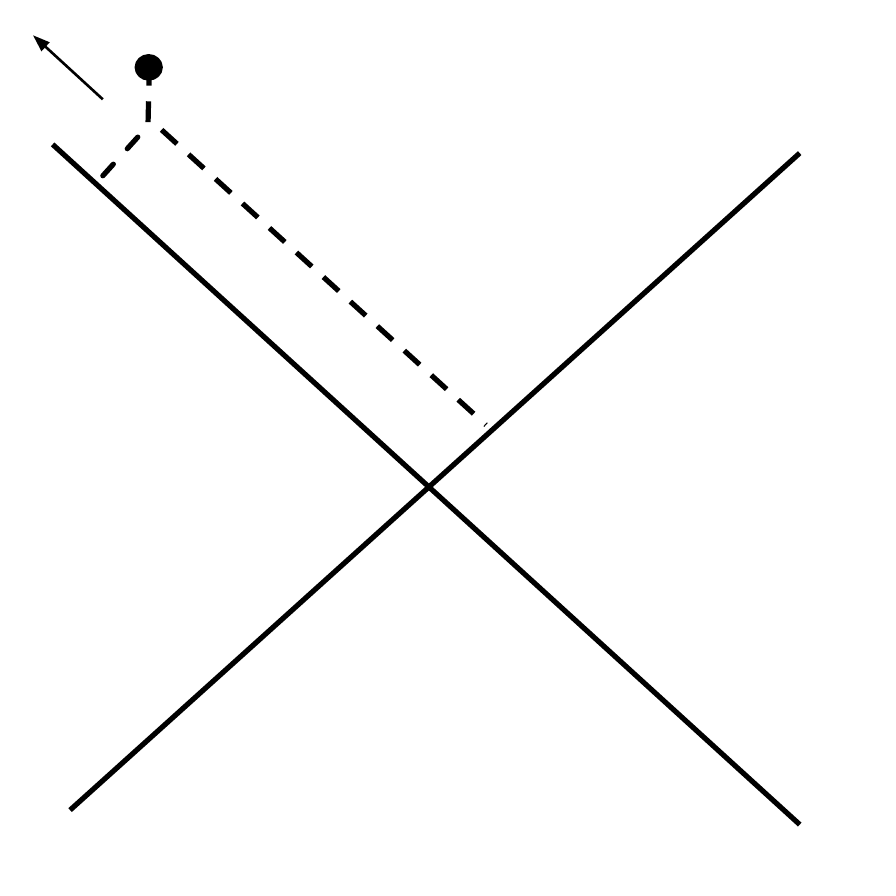}
\caption{Approaching a pole on $\p\Sigma$ corresponds not only to movement in the transverse direction towards the surface of the external brane corresponding to the pole, but also to a longitudinal movement along the brane off to infinity (in the direction of the arrow). In cases where the probe string must couple to two branes at once, approaching a pole always leads to one of the open strings stretching to infinite length.}
\label{fig:divergence}
\end{figure}

Having explained the mechanism behind the vanishing of probe energies at certain poles, we now turn towards the explanation of the divergences in energies at all the remaining poles. We begin by recalling the nature of the near-pole expansion given in (\ref{sfmnp}) - as the radial distance $r$ from the pole is brought to zero, the $AdS_6$ turns into six-dimensional Minkowski space $\RR^{1,5}$. The natural interpretation of this in the brane web picture is that as the pole is approached,  one moves not only along the transverse direction towards the surface of the $(p,q)$ five-brane (as mentioned above), but also simultaneously along the longitudinal direction out to infinite distance from the intersection point of the web, where the geometry becomes the near-horizon geometry of a single infinite $(p,q)$ five-brane.\footnote{Further evidence for this interpretation is obtained by noting that the metric factor $\tilde{\rho}^2$ diverges at the poles, so that each pole has infinite proper distance from any other point in $\Sigma$. I thank Christoph Uhlemann for this point.} This is illustrated in Figure \ref{fig:divergence}. Thus for probe strings which cannot interact directly with the brane they are approaching, there will always be at least one open string which is stretched to infinite length as one moves towards the pole. This explains the divergences in Figures \ref{fig:4sym}, \ref{fig:4anti},  \ref{fig:4}, \ref{DString}.

We should repeat that in general, the radial distance of the probe away from the intersection point has no meaning in the supergravity solutions - at the end of the day, the probe should really be taken to be localized at the intersection point. The exception is when we approach the surface of a brane, while simultaneously moving longitudinally along the brane out towards infinity. This has an interpretation in the supergravity solutions as a motion towards the corresponding pole on the boundary of $\Sigma$.

 The angular position of the probe around the intersection point also has an interpretation on the supergravity side. When $\Sigma$ is taken to be the unit disc, the angular coordinate in the plane of the web roughly corresponds to the angular coordinate along $\p\Sigma$. Concretely, the various quadrants (or more generally ``sectors") of the brane web correspond to arcs along the unit circle, bounded on either side by the poles corresponding to the external branes bounding the sector. This is the reason for the claim that $\Sigma$ can be thought of as being ``located" at the intersection point. By a simple conformal transformation, the correspondence between sectors of the web and arcs of the unit circle can be rephrased as a correspondence between sectors of the web and intervals on the real line for $\Sigma$ the upper-half plane. We will see further details of this correspondence momentarily when we explore the three-pole case.

\begin{figure}
\centering
\includegraphics[scale=0.6]{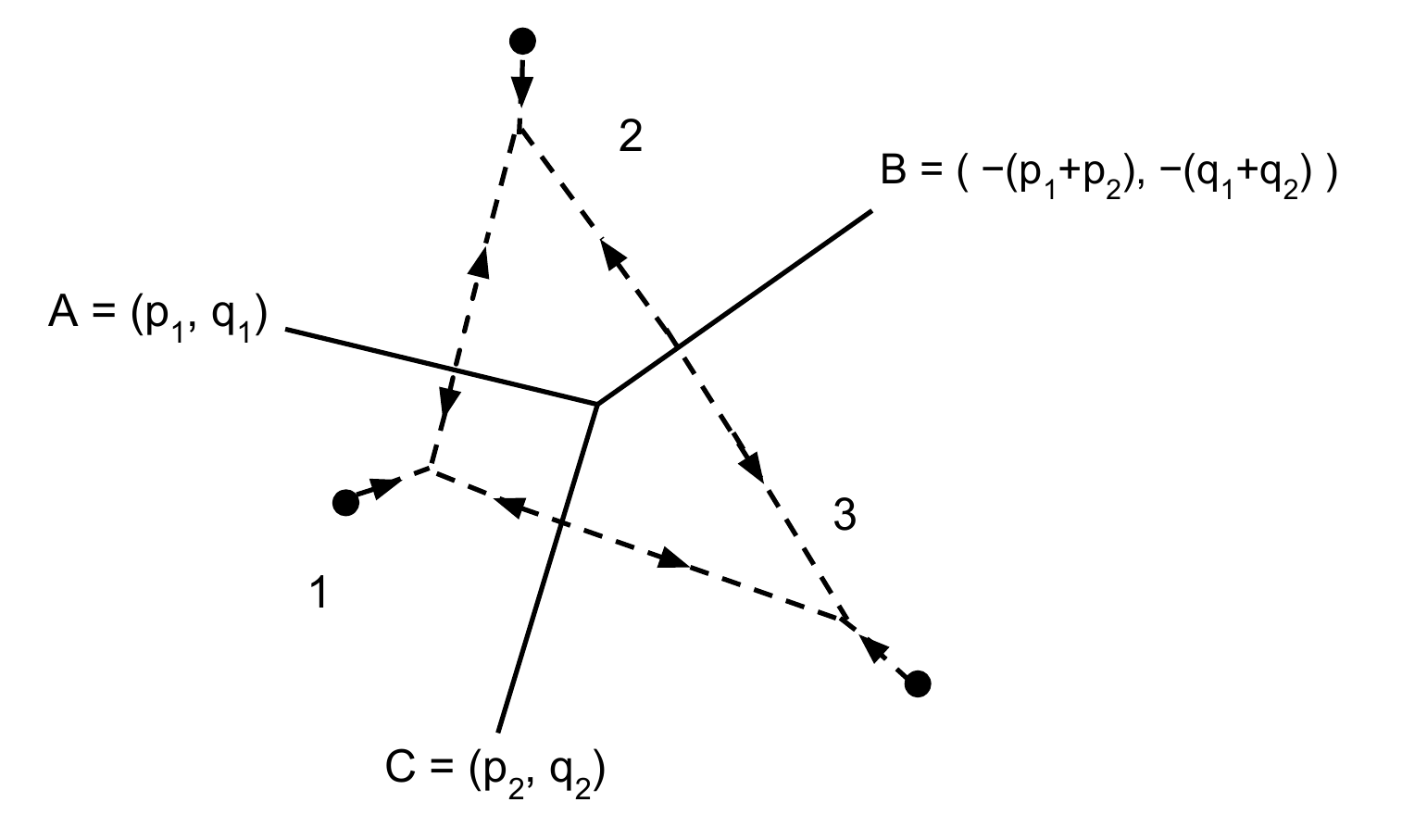}
\caption{There are three sectors of junctions allowed in the generic three-pole case. }
\label{3polejunction}
\end{figure}

Due to its symmetry, the four-pole case analyzed above is actually simpler than the generic three-pole case given by (\ref{3poledata}). As shown in Figure \ref{3polejunction}, in the three-pole case there are three sectors, each of which gives rise to distinct junctions. The probe strings which can form junctions in each sector are (integer multiples of): 
\begin{itemize}
\item{ \textbf{Sector 1:} $(q_1 - q_2\,, p_2 - p_1)$-strings}
\item{ \textbf{Sector 2:} $\left(-(2 q_1 + q_2)\,, 2 p_1 + p_2 \,\right)$-strings}
\item{ \textbf{Sector 3:} $\left( q_1 + 2 q_2, -( p_1 +2  p_2) \,\right)$-strings}
\end{itemize}

\begin{figure}
\centering
\begin{minipage}{.5 \textwidth} 
\centering
\includegraphics[scale=0.83]{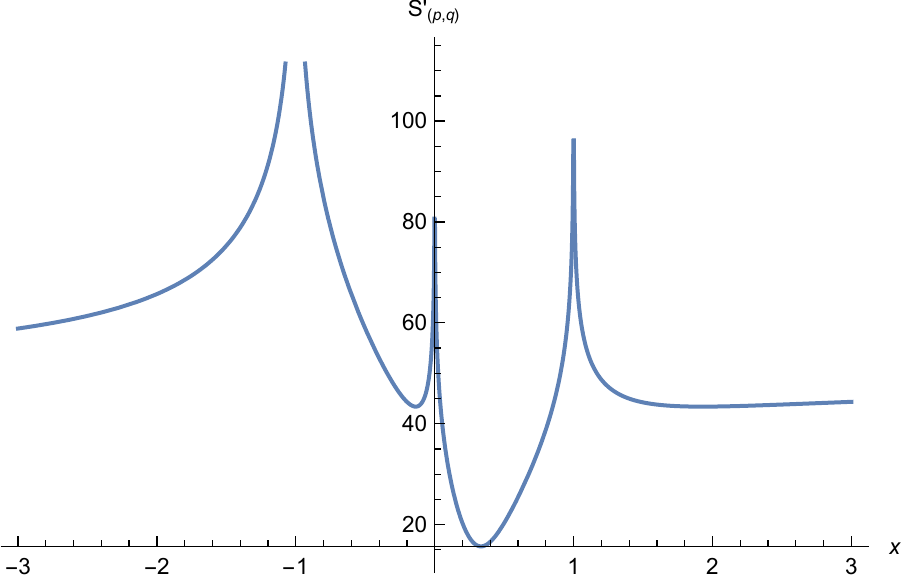}
\end{minipage}%
\begin{minipage}{.5 \textwidth} 
\centering
\includegraphics[scale=0.83]{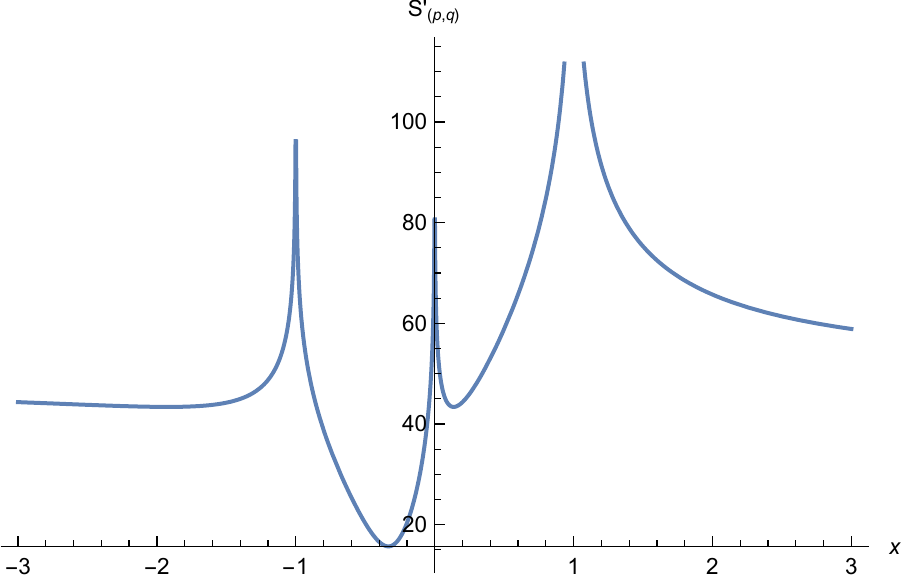}
\end{minipage}%
\caption{The energy profiles of Sector 2 probe strings (left) and Sector 3 probe strings (right). The locations of the global minima can be understood by considering which two external branes the probe string couples to. For explicit results, we have taken $(p_1, p_2, q_1, q_2) = (1, 1, -1, 1)$. }
\label{3poleplots}
\end{figure}

As in the four-pole cases above, one can check that the energy profiles of probe $(p,q)$-strings with the above charges are independent (up to constant factors) of the choice of background charges, since they always form junctions of the same type. In particular, the energy profiles of the probe strings in Sectors 1, 2, and 3 are always of the form shown in Figures \ref{fig:4}, \ref{3poleplots}a, and \ref{3poleplots}b, respectively. With the string junction picture in mind, we now aim to explain the general shape of these plots. First recall that as per (\ref{3poledata}), the $(p_1,q_1)$-brane is located at $x=1$, the $\left( -(p_1 +p_2)\,,\,-(q_1 + q_2)\,\right)$-brane is at $x=0$, and the $(p_2,q_2)$-brane is at $x=-1$ on the boundary of $\Sigma$. For simplicity, from now on we refer to these branes  as A-, B-, and C-branes respectively. As in Figure \ref{3polejunction}, the Sector 2 probe string forms junctions with open strings anchored on the A- and B-branes, and hence via the correspondence between sectors and boundary intervals the Sector 2 probe string should locate itself somewhere between $x=1$ and $x=0$ on the real line. This explains the rough location of the global minimum in Figure \ref{3poleplots}a. Likewise, the Sector 3 probe string forms junctions with open strings anchored on the B- and C-branes - thus it should be located somewhere between $x=0$ and $x=-1$. This explains the behavior in Figure \ref{3poleplots}b. Finally, the Sector 1 probe string forms junctions with open strings anchored on the A- and B-branes, and so should be located between $x=1$ and $x=-1$, i.e. at the point at infinity (this is clearer when $\Sigma$ is thought of as a unit disc, where the point at infinity is just a usual point on the boundary of the disc). This explains the location of the global minimum of Figure \ref{fig:4}. All of this same logic may be applied to explaining the location of the global minima of the energy profiles in the four-pole cases as well. 

\begin{figure}
\centering
\begin{minipage}{.5 \textwidth} 
\centering
\includegraphics[scale=0.83]{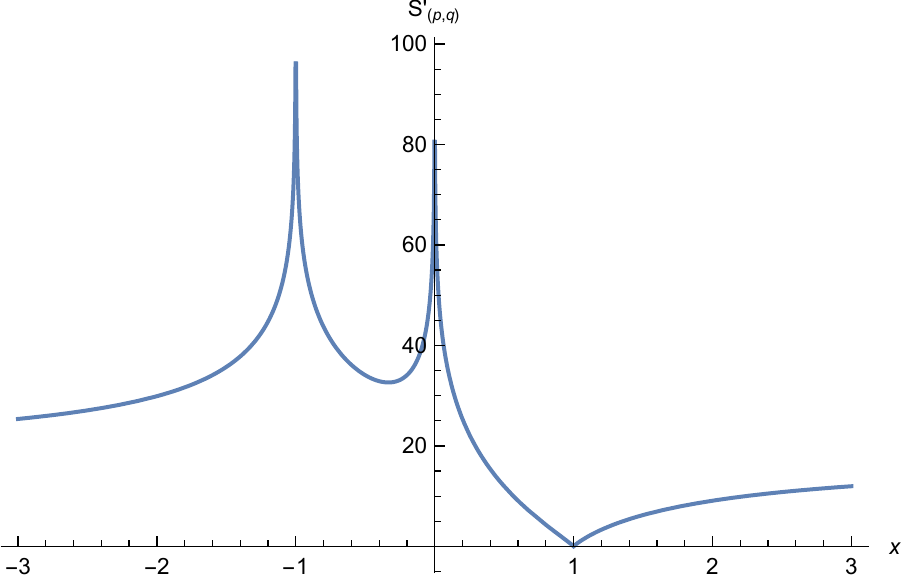}
\end{minipage}%
\begin{minipage}{.5 \textwidth} 
\centering
\includegraphics[scale=0.83]{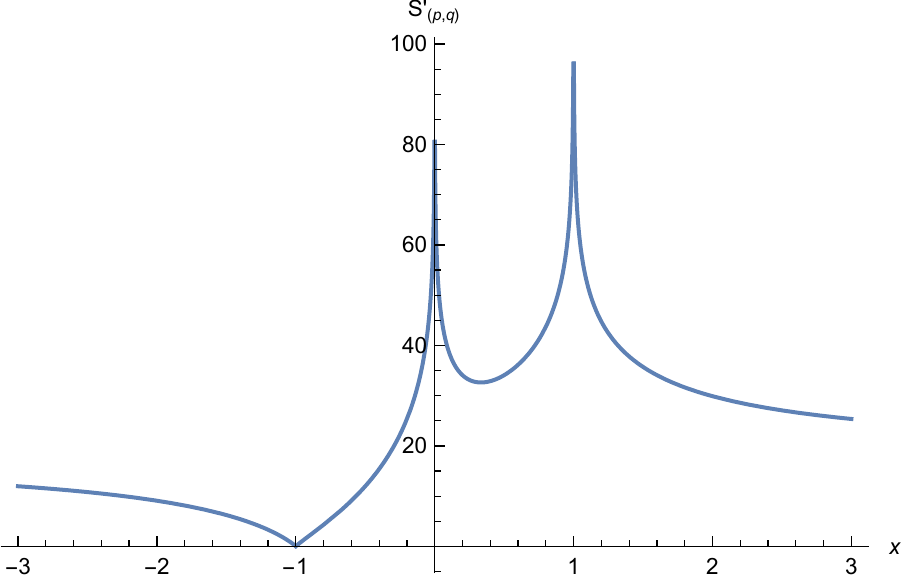}
\end{minipage}%
\newline
\begin{minipage}{.45 \textwidth} 
\centering
\includegraphics[scale=0.83]{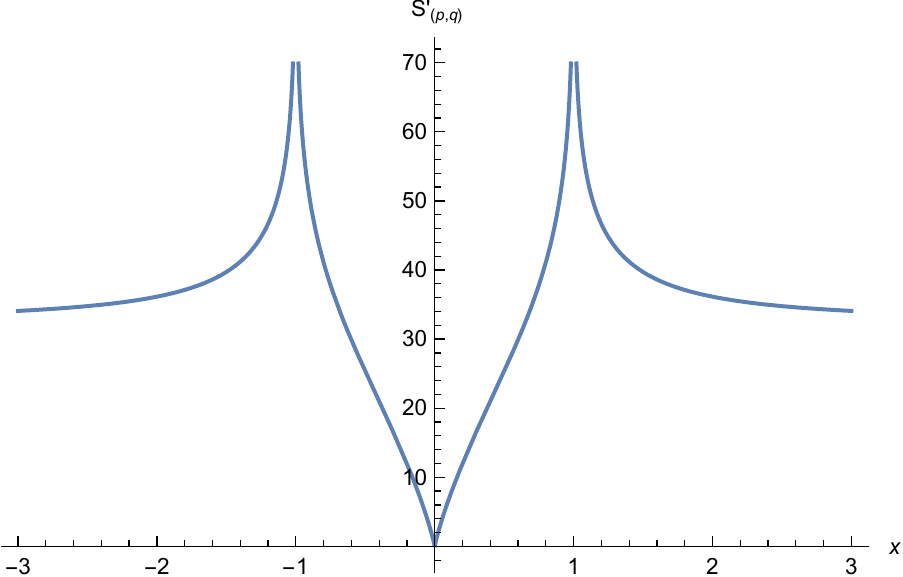}
\end{minipage}%
\caption{Energy profiles of probe string embeddings in three-pole backgrounds (\ref{3poledata}). The probe string is taken to have the following charges: top row, left to right: $\pm (- q_1, p_1)$ and $\pm (- q_2, p_2)$, bottom row:  $\pm\left( q_1 + q_2 , -( p_1 +  p_2) \,\right)$. These have vanishing energy when they lie on A-, C-, and B- branes, respectively. For explicit results, we have taken $(p_1, p_2, q_1, q_2) = (1, 1, -1, 1)$. }
\label{3poleonbrane}
\end{figure}

Finally, we may check the cases of probe strings of charges $\pm (- q_1, p_1)$, $\pm (- q_2, p_2)$, and $\pm\left( q_1 + q_2 , -( p_1 +  p_2) \,\right)$. Since these can couple directly to a single brane of the A-, B-, or C-type respectively, they are expected to have vanishing energy when they lie directly on those branes. Indeed, this is what is observed in Figure \ref{3poleonbrane}. So by purely microscopic, brane web considerations, we have managed to predict some of the qualitative aspects of the probe string energy profiles.

\subsection{String Tension in Supergravity}

We now explore more quantitative aspects of the intuition above. To begin, recall that in Section 3, minimal energy F1-string embeddings were found to occur when $\mathrm{Re}\,\f_+' = 0$. These embeddings were supersymmetric (i.e. 1/2-BPS) when $\mathrm{Im}\,\f_+ = 0$ as well. Similar results were found for D1-strings in Section 4, namely that minimal energy occurred when $\mathrm{Im}\,\f_+' = 0$ and supersymmetry was ensured for $\mathrm{Re}\,\f_+ = 0$.  We would now like to derive these results via microscopic considerations. In particular, the minimal energy embeddings should occur when the total tension of open strings in the junction is minimized. 

For concreteness, we focus on backgrounds which admit F1- and D1-string embeddings supported by only a single junction, as opposed to multi-junction configurations to be discussed briefly later. For four-pole cases given by (\ref{harderembedding}), this amounts to the requirement that $p_1 = p_2$ for F1-strings and $q_1 = q_2$ for D1-strings, since in these cases $(q_2 \pm q_1, p_2 \mp p_1)$ has one vanishing entry. We will only be concerned with embeddings which are supersymmetric, and so by the considerations of Sections 3.3 and 4 we must restrict to backgrounds with \textit{both} $p_1 = p_2$ and $q_1=q_2$. With this in mind, we make the following conjecture:
\newline
\newline
\textbf{Strong Conjecture:} For four-pole backgrounds (\ref{harderembedding}) satisfying $p_1 = p_2$ and $q_1 = q_2$, the quantity $K(x) \equiv (q_1 \mathrm{Re}\,\f_+)^2 + (p_1 \mathrm{Im}\,\f_+)^2 $ can be interpreted as the total tension of the junction formed between the probe F1/D1-strings and open strings anchored on the branes of the web. More precisely, we have the equality
\bea
\label{conjecture}
T(x) = {c_1 \over (p_1 q_1)^{3/2}}\, K(x) + \sqrt{p_1 q_1}\, c_2
\eea
where $T(x)$ is the total junction tension at the point $x$ and $c_1$, $c_2$ are background independent constants.
\newline
\newline
The factors of $(p_1 q_1)$ have been included to ensure that both sides of (\ref{conjecture}) transform in the same way under separate scalings $p_1 \rightarrow \lambda_p \,p_1$ and $q_1 \rightarrow \lambda_q\, q_1$. With this identification, we will see that the minimal energy conditions above become simple tension extremization conditions. Note that in the simplest, symmetric background (\ref{easyembedding}), the function $K(x)$ reduces to $N^2 \,| \f_+(x)|^2$. Up to branch cut contributions, this is none other than the norm-square of the boundary value of the defining supergravity functions $\cA_+$; see (\ref{lastone}).

\begin{figure}
\centering
\centering
\includegraphics[scale=0.9]{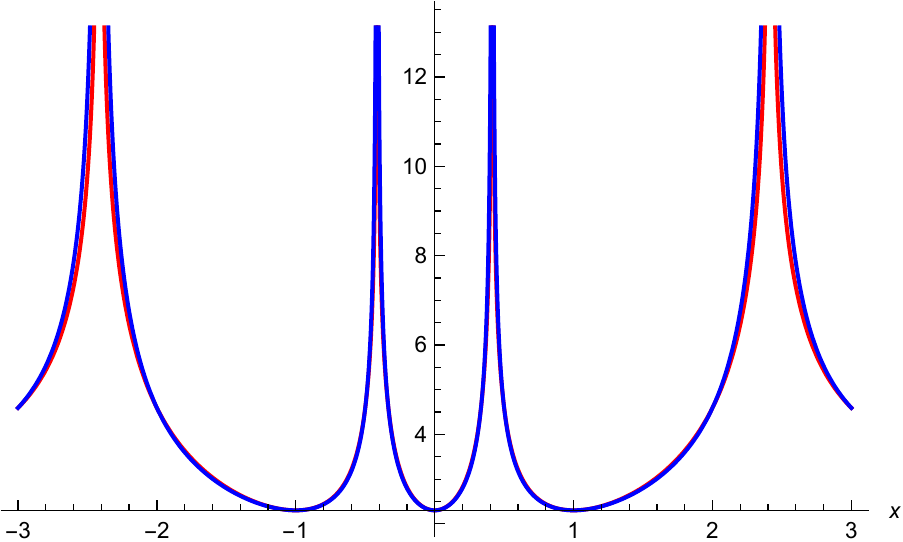}
\caption{The function ${c_1\over (p_1 q_1)^{5/2}} K(x) +c_2$ for a four-pole background with $p_1 = p_2$ and $q_1 = q_2$ is plotted in red. The total junction tension $T(x)/\sqrt{p_1 q_1}$ for the same background is plotted in blue. The only slight deviations between the two are near the poles, where divergences occur and the plots are not expected to be reliable anyways.}
\label{energyprofile}
\end{figure}

To justify the identification of the right-hand side of (\ref{conjecture})  with the total tension, we begin by plotting the former. This is shown in Figure \ref{energyprofile}. We have divided by a factor of $ \sqrt{p_1 q_1}$ so that the result is unchanged for any choice of $p_1$ and $q_1$. 

 We now aim to recreate this profile by calculating the tension of the string junction formed between the brane web and the probe string. The image to keep in mind is that of Figure \ref{fig:branejunc}. In that case, both of the junctions involved open strings with charges $\pm (-q_1,p_1)$ and $\pm (q_2,p_2)$, with the respective tensions $T_1$, $T_2$ given by
\bea
\label{tensions}
T_1 = e^\phi \sqrt{p_1^2 \, e^{-4 \phi} + (q_1 + p_1 \chi)^2} \hspace{0.7 in} T_2 = e^\phi \sqrt{p_2^2 \, e^{-4 \phi} + (q_2 - p_2 \chi)^2}
\eea
as can be read off by comparing the $(p,q)$-string action to the action of the F1-string \cite{Schwarz:1995dk}. Depending on which of the two sectors we are in, the third open string in the junction is either of type $- (q_2 + q_1, p_2 - p_1)$ or of type  $- (q_2 - q_1, p_2 + p_1)$, and the tensions are the appropriate generalizations of the expressions above. 

 All of the same considerations hold in the current case, where $p_1 = p_2$ and $q_1 = q_2$. In this case the third open strings in the junction are either bound states of $2 q_1$ F1-strings or bound states of $2 p_1$ D1-strings.  

The tensions above are all functions of location on $\Sigma$. The natural location $x$ at which to evaluate these tensions is the $x$ at which the probe string is located. A simple check of this choice of $x$ is to plot these tensions as functions of the location of the probe - as expected, one finds that the tension of a given open string vanishes precisely when the probe is directly atop the brane from which that open string originated. 

Now we may define the total junction tension $T(x)$. We take the total tension of a junction to be the simple sum of the tensions of the three strings in the junction. For the junctions containing F1-strings, this is a sum of the tensions $T_1$ and $T_2$ of (\ref{tensions}) together with the tension of $2 q_1$ F1-strings. These junctions form in the upper and lower quadrants of the brane web, which in terms of $\Sigma$ translate to the boundary regions $-1-\sqrt{2} \leq x \leq 1- \sqrt{2}$ and $-1+\sqrt{2} \leq x \leq 1+ \sqrt{2}$ on the real line via the sector-interval correspondence.  For the junction containing a D1-string, the total tension is a sum of the tensions $T_1$ and $T_2$ with the tension of $2 p_1$ D1-strings. These junctions form in the left and right quadrants of the brane web, which are the remaining boundary regions. Thus we have the following piecewise form for the total tension $T(x)$, 

\bea
T(x) = \left\{ 
\begin{array}{l}
T_1(x) + T_2(x) + |2 q_1|\, T_F(x) \hspace{0.5 in} -1-\sqrt{2} \leq x \leq 1- \sqrt{2}\\
T_1(x) + T_2(x) + |2 q_1|\, T_F(x) \hspace{0.5 in}  -1+\sqrt{2} \leq x \leq 1+ \sqrt{2}\\
T_1(x) + T_2(x) + |2 p_1| \,T_D(x) \hspace{0.5 in} \qquad\qquad\mathrm{otherwise}
\end{array}
\right.
\eea

In Figure \ref{energyprofile} we have plotted ${T(x)\over  \sqrt{p_1 q_1}}$ together with the function $ {c_1 \over (p_1 q_1)^{5/2}}\, K(x) + c_2$ in a four-pole background with $p_1 = p_2$ and $q_1 = q_2$. The constants $c_1$ and $c_2$ were chosen to get a match between the two. Their numerical values are  
\bea
\label{constsAB}
c_1 =0.43733\ldots \hspace{1.5 in} c_2 = 3.24698\ldots
\eea
though closed form expressions for them remain unknown. We see that with this choice of constants, the two plots are nearly identical. The only points of discrepancy are those in close vicinity to the poles, where the plot is not expected to be reliable anyways. Thus we have numerical evidence in support of the conjecture. Actually though, in what follows we will require only a weaker form of the conjecture, 
\newline
\newline
\textbf{Weak Conjecture:} For four-pole backgrounds satisfying $p_1 = p_2$ and $q_1 = q_2$, the functions $T(x)$ and $ {c_1 \over (p_1 q_1)^{3/2}}\, K(x) + \sqrt{p_1 q_1}\, c_2$ have the same locations and values of all of their minima. 
\newline
\newline
This weaker conjecture has been verified to hold for all choices of $p_1$ and $q_1$ between $1$ and $10$. Thus even if $K(x)$ is not  equivalent to $T(x)$ in the sense of (\ref{conjecture}), we will assume that the two yield identical minimization problems.

Assuming at least this weaker conjecture to hold, we can now explain the energy minimization conditions obtained for the F1- and D1-strings on the supergravity side via microscopic considerations. In particular, since the linear function of $K(x)$ on the right-hand side of (\ref{conjecture}) has the same minima as the total junction tension, one may minimize tension by just minimizing the function $K(x)$, 
\bea
\p_x K(x) =2  \left[ (q_1)^2 \, \mathrm{Re}\, \f_+  \,\mathrm{Re}\, \f'_+ + (p_1)^2 \, \mathrm{Im}\, \f_+ \, \mathrm{Im}\, \f'_+ \right]= 0
\eea
Since we do not want $\f_+$ or $\f'_+$ to vanish identically, nor can we have $p_1 = p_2 = 0$ or $q_1 = q_1 = 0$ to have a legitimate brane web,  there are only two ways to satisfy this equation:
\bea
  \mathrm{Re}\, \f'_+=0  \hspace{0.1 in} \mathrm{and} \hspace{0.1 in}\mathrm{Im}\, \f_+ =0 \hspace{0.5 in} \mathrm{or}  \hspace{0.5 in}   \mathrm{Im}\, \f'_+=0  \hspace{0.1 in} \mathrm{and} \hspace{0.1 in}\mathrm{Re}\, \f_+ =0
  \no
  \eea
 Only the first conditions have solutions in the regions $-1\pm\sqrt{2} \leq x \leq 1\pm \sqrt{2}$, while only the second have solutions in the remaining regions. These are exactly the conditions expected for the F1- and D1-strings, respectively! Note that not only do we reproduce the minimal energy condition of the supergravity side, but also the supersymmetry condition. Thus using our conjecture we have managed to recast the minimization problem on the supergravity side as a minimization problem on the brane web side.

\begin{figure}
\centering
\includegraphics[scale=0.6]{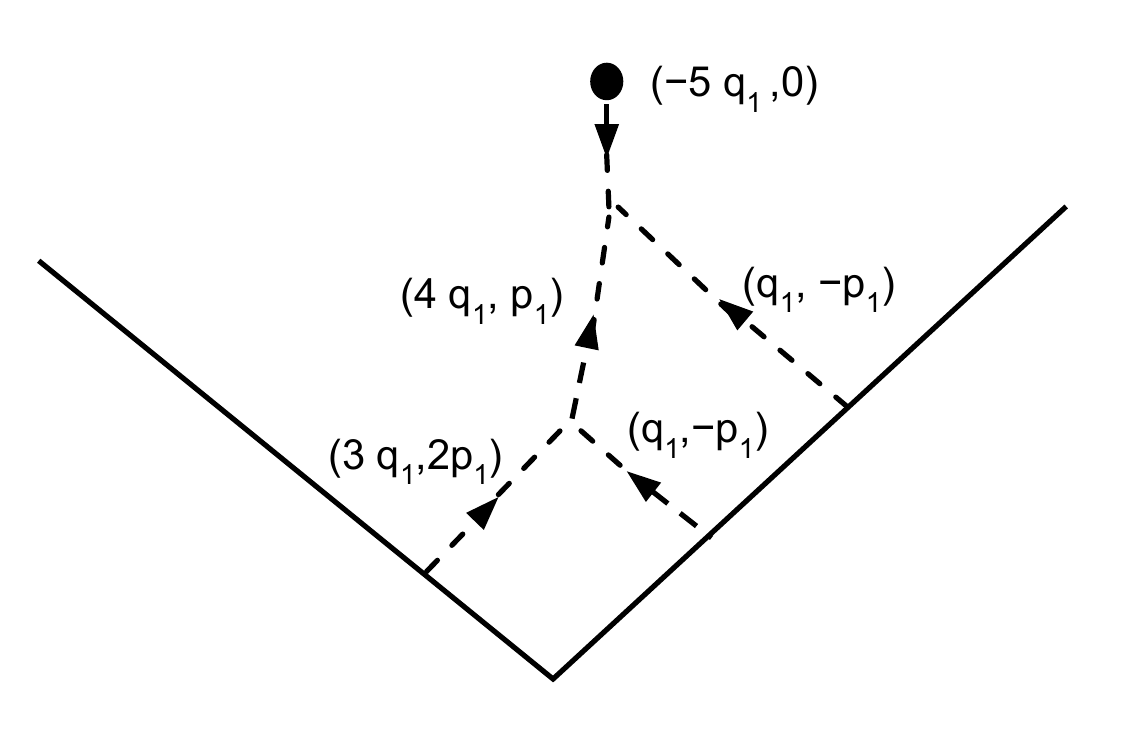}
\caption{An example of a multi-junction configuration for the four-pole background of (\ref{harderembedding}) with $p_2 = 2 \,p_1$ and $q_2 = 3\, q_1$.}
\label{susynonminen}
\end{figure}
One may now move on to the other backgrounds allowing supersymmetric minimal energy embeddings, such as the case with $p_2 = 2 \,p_1$ and $q_2 = 3\, q_1$.  However, if we now try to examine the interactions of probe F1- or D1-strings with the background web, by charge conservation we would necessarily have to consider multi-junction configurations. An example of a multi-junction configuration is shown in Figure \ref{susynonminen}. In the particular case shown, the total junction tension $T(x)$ is the sum of the tensions of the five open strings involved in the junction. Unfortunately, it is not yet clear what quantity on the supergravity side this tension would correspond to, though one could hope that it is some simple generalization of the function $K(x)$ above. We save the identification of this more general quantity for a promising future project.

There is a final point that should be addressed. So far we have only looked at interactions between the probe string and the web mediated by open strings stretching between the two. But in general there are also interactions due to closed strings sourced by the branes. It is these closed strings which change the background geometry in which our probe string is placed, as well as generate a non-trivial dilation profile. Fortunately though, these closed strings should not affect any of the considerations of this section - after all, closed strings do not couple to the $(p,q)$-charges we have been studying here.

\section{Discussion}
\setcounter{equation}{0}
In this paper, we explored a variety of probe string embeddings in supergravity backgrounds with geometry $AdS_6 \times S^2$ warped over a Riemann surface $\Sigma$. In doing so, we found significant numerical evidence in support of the interpretation of these backgrounds as the near-horizon geometry of $(p,q)$ five-brane webs in the conformal limit. Moreover, we saw that a certain quantity $K(x)$ on the supergravity side can be given a microscopic interpretation as string tension via the brane web picture. Unfortunately, this identification is restricted to the very special case of four-pole backgrounds (\ref{harderembedding}) with $p_1 = p_2$ and $q_1 = q_2$. In the even more symmetric case where all of the charges are equivalent (up to signs), the total open string tension is in fact given by $|\f_\pm|^2$, which we see from (\ref{lastone}) is nearly the norm-square of the boundary value of the defining supergravity functions $\cA_\pm$, with branch cut contributions neglected. It remains to be seen how these results generalize to more complicated backgrounds. Since our microscopic considerations are most illuminating in cases which preserve supersymmetry, the next obvious target would be the  $(p_2, q_2) = (2 p_1, 3 q_1)$ configuration. As mentioned before, this would involve multiple three-string junctions. 

Another direction of generalization is to relax the restriction made in this paper of considering only probe string embeddings which preserve the maximal number of bosonic symmetries, i.e. all of the $SO(2,1) \times SO(4) \times SU(2)_R$. Breaking some of these bosonic symmetries usually breaks more (or all) of the background supersymmetries too. However, allowing for breaking of the $SO(4)$ symmetry does not seem to do much. In terms of superalgebras, breaking of the $SO(4)$ can be accomplished by breaking only the $A_1$ part of $A_1 \oplus D(2,1;2)$, thus leaving the superalgebra $D(2,1;2)$ and its 8 supercharges intact. In terms of the string action, relaxing the condition of $SO(4)$ symmetry just amounts to not enforcing $u = 0$ for the location of the probe string embedding. Then when one proceeds to calculate the Nambu-Goto action for the string, one finds that the only $u$-dependent contribution is in the form of an overall multiplicative factor of $\cosh u $. Thus extremizing the action with respect to $u$ sets $u = 0$ anyways, restoring the $SO(4)$. 

On the other hand, configurations breaking the other two symmetries are significantly more difficult to work with. In particular, the only reason we were able to make reasonable analytic progress in Sections 3 and 4 was because we were locating ourselves on the boundary of $\Sigma$, where many of the supergravity quantities had simplifying expansions. If we were to try to wander away from the boundary of $\Sigma$, thus breaking the $SU(2)_R$ symmetry, we would run into a significant amount of computational difficulty. There is also an interpretational difficulty here, as it is not entirely clear what moving into the interior of $\Sigma$ even corresponds to in the brane web picture.\footnote{However, there is a reasonable guess for the significance of the interior of $\Sigma$. Namely, points in the interior of $\Sigma$ may correspond to locations within the internal structure/faces of the web. This idea has been advanced in \cite{DHoker:2017zwj} for the case of 7-branes inserted in the web. With this interpretation, one may actually hope to reproduce the BPS states found in \cite{Aharony:1997bh, Kol:1998cf} by moving the probe strings studied here to the interior of $\Sigma$.} This would be an interesting avenue of future research.

Finally, we close with some words about the interpretation our embeddings in the dual five-dimensional SCFT. In \cite{Aharony:1997bh, Kol:1998cf}, open strings and junctions anchored on internal branes were given an interpretation as BPS states in the field theory living on those internal branes. For example, an F1-string stretched between internal D5-branes corresponds to a massive gauge boson, while a D1-string stretched between internal NS5-branes corresponds to an instanton. As already mentioned above though, the junctions in those cases are of a fundamentally different nature than those discussed here, since in the current case the open strings of the junction are anchored on \textit{external} $(p,q)$ five-branes. Thus our junctions do not possess the same interpretation as BPS particles in the field theory. However, the probe string itself does have a possible interpretation as a 1/2-BPS Wilson - `t Hooft loop in the five-dimensional SCFT. As is well known, Wilson loops in the fundamental representation can be calculated holographically via minimal-area F1-strings anchored along the contour of the Wilson loop on the boundary and extending into the bulk \cite{Maldacena:1998im,Drukker:1999zq}.  More general Wilson - `t Hooft loops can be obtained in a similar way by embedding probe $(p,q)$-strings instead of probe F1-strings \cite{Chen:2006iu}, though we will focus for simplicity on pure Wilson loops below.  

The probe F1-string embeddings explored in Section 3 are exactly those that would be dual to 1/2-BPS, circular Wilson loops on the boundary of $AdS_6$ - indeed, the presence of the circular loop would break the $SO(5, 2)$ conformal symmetry of the boundary to an $SO(2,1)$ conformal symmetry and a transverse $SO(4)$, the same symmetries as for our probe string embedding. In fact, the results of Section 3 already provide the necessary ingredients for the computation of expectation values of such loops via the usual prescription,
\bea
\langle W_{fund}\rangle \sim e^{\,S_{F1}'} \sim e^{\, \tilde{f}_6^2}
\eea
where the $\sim$ indicates equivalence up to an overall multiplicative factor and we have used  (\ref{needforwl}). The function $\tilde{f}_6^2$ is to be evaluated at an extremum on the boundary of $\Sigma$. From Section 3.1, we know that minimal energy embeddings are obtained by solving $\mathrm{Re}\,\f'_+ =0$ to get some solutions $x_\pm$ on the boundary. For those backgrounds which admit supersymmetric minimal energy embeddings, we have by (\ref{simplesff62}) that at these extremal points the renormalized action takes the form $\tilde{f}_6^2(x_\pm) = 4 \sqrt{2}\, |\f_+(x_\pm)|$. 

To be completely explicit, let's evaluate this for the usual four-pole case of (\ref{harderembedding}). In this case the $x_\pm$ are given by (\ref{4polemin}) and one finds that 
\bea
|\f_+(x_\pm)| = \left| p_2 \log{\left({\mp p_1 + \sqrt{p_1^2 + p_2^2} \over p_2 }\right)}  - p_1\log{\left({\pm p_2 + \sqrt{p_1^2 + p_2^2} \over p_1 }\right)}  \right|
\eea
A simple calculation shows that $\f_+(x_+) = - \f_+(x_-)$, so that the two solutions $x_\pm$ give the same results for the Wilson loop expectation values. Then for example if $p_1 = p_2$, the Wilson loop expectation value is given by
\bea
 \langle W_{fund} \rangle \sim e^{4 \sqrt{2} \,p_1 \log{\left[3 \,+\, 2\sqrt{2}\right]}}
 \eea
 
While all of the necessary computational tools are certainly present here, there is some interpretational difficulty with these Wilson loops. This is because the 5d SCFTs expected to be dual to these supergravity solutions are generically quiver theories \cite{DHoker:2017mds,Bergman:2012kr}, in which case it is unclear what one even means by ``fundamental representation". Though such quiver theory Wilson loops have been studied in the past, see e.g. \cite{Rey:2008bh,Hatsuda:2013yua,Bigazzi:2008zt} and also the especially relevant \cite{Assel:2012nf} (which explores a similar 5d example), we do not provide any further comments on them here, instead relegating them to future works. 
\newline
\newline
\newline
\newline
\newline
{\Large\textbf{Acknowledgements}}
\newline
\newline
The author would like to thank Michael Gutperle, Christoph Uhlemann, and especially Eric D'Hoker for their invaluable guidance and much imparted knowledge. He would also like to thank Himanshu Raj for many useful conversations, as well as the speakers and organizers of TASI 2017, where part of this work was completed. Finally, he would like to thank A.K.D.
\newpage
\appendix 
\section{Gamma Matrix Conventions}
\setcounter{equation}{0}
For ease of reference, throughout this paper we use the same gamma matrix conventions as in \cite{DHoker:2016ujz}. We reproduce these conventions here for completeness. The full ten-dimensional Dirac matrices satisfy $\{ \Gamma^M, \Gamma^N \} = 2 \eta^{MN} \mathds{1}_{32}$ and take the form 
\bea
\Gamma^m &=& \g^m \otimes \mathds{1}_2  \otimes \mathds{1}_2 \hspace{1 in} m = 0,1,2,3,4,5
\no\\
\Gamma^i &=& \g_{(1)} \otimes \g^i  \otimes \mathds{1}_2 \hspace{1 in} i \,= \,6,7
\no\\
\Gamma^a &=& \g_{(1)} \otimes \g_{(2)}  \otimes \g^a \hspace{0.9 in} a \,= \,8,9\hspace{0.3 in}
\eea 
while the lower-dimensional gamma matrices are taken to be 
\bea
\g^0 &=& - i \, \s^2  \otimes \mathds{1}_2\otimes \mathds{1}_2
\no\\
\g^1 &=& \s^1 \otimes \mathds{1}_2\otimes \mathds{1}_2
\no\\
\g^2 &=& \s^3 \otimes \s^2 \otimes \mathds{1}_2 \hspace{0.9 in} \g^6 = \s^1
\no\\
\g^3 &=& \s^3 \otimes \s^1 \otimes \mathds{1}_2 \hspace{0.9 in} \g^7 = \s^2
\no\\
\g^4 &=& \s^3 \otimes \s^3 \otimes \s^1 \hspace{0.9 in} \g^8 = \s^1
\no\\
\g^5 &=& \s^3 \otimes \s^3 \otimes \s^2 \hspace{0.9 in} \g^9 = \s^2
\eea
and they satisfy $\{\g^{m} , \g^{n}\} = 2\, \eta^{m n}$, $\,\{\g^{i} , \g^{j}\} = 2\, \delta^{i j}$, and $\{\g^{a} , \g^{b}\} = 2 \,\delta^{a b}$.  Above, we have also made use of the following chirality matrices, 
\bea
\g_{(1)} = \s^3 \otimes \s^3 \otimes \s^3 \hspace{0.8 in} \g_{(2)} = \s^3 \hspace{0.8 in} \g_{(3)} = \s^3 
\eea
Finally, we will need an explicit form of the complex conjugation matrices,
\bea
\label{compconmat}
\left(\g^m\right)^* = + B_{(1)} \g^m B_{(1)}^{-1} \hspace{1.05 in} B_{(1)} &=& - i \g^2 \g^5 = \mathds{1}_2 \otimes \s_1 \otimes \s_2
\no\\
\left(\g^i\right)^* = - B_{(2)} \g^i B_{(2)}^{-1} \hspace{1.1 in}B_{(2)} &=& \g^7 = \s^2
\no\\
\left(\g^a\right)^* = - B_{(3)} \g^a B_{(3)}^{-1} \hspace{1.08 in}B_{(3)} &=& \g^9 = \s^2
\eea
\newpage
\section{Boundary Expansions}
\setcounter{equation}{0}

In this appendix, we give the boundary expansions of various supergravity fields. Most of these expansions were first given in \cite{Gutperle:2017}. We begin by repeating some of their results, and then continue by deriving further ones which will be of use to us in this paper. 

As in the main text, we take the Riemann surface (with boundary) $\Sigma$ to be the upper-half plane with local complex coordinates $w, \bar w$ and its boundary to be the real axis. As usual, the coordinate along the real axis is denoted $x$, while that along the imaginary axis is denoted $y$. Then a near-boundary expansion amounts to a power series expansion in $(i y)$, and we have for example 
\bea
\label{lastone}
\cA_\pm(w) = D_\pm(x) + \sum_{n = 0}^\infty {1 \over n!} (i y)^n \f_\pm^{(n)}(x)
\eea
where 
\bea
D_\pm(x) = i \pi \sum_{\ell=1}^L Z_\pm^\ell \Theta\left(r_\ell - x \right)  \hspace{0.5 in} \mathrm{and}\hspace{0.5 in}  \f_\pm(x) = \cA^0_\pm + \sum_{\ell = 1}^L Z_\pm^\ell \log|x - r_\ell|
\eea
The $D_\pm$ arise due to the logarithmic branch cuts along the real axis, which are taken to run off in the negative $x$-direction. Note that in the gauge $\cA_+^0 + \bar \cA_-^0 = 0$, we have $\f_\pm = - \bar \f_\mp$. The derivatives of $\cA_\pm$ are easily evaluated, 
\bea
\k_\pm \equiv \p_w \cA_\pm = \sum_{n=0}^\infty {1\over n!} (i y)^n \f_\pm^{(n+1)}
\eea
From this, an expression for $\k^2 = - |\k_+|^2 +|\k_-|^2 $ is obtained, 
\bea
\k^2 = y \, \k_{(1)}^2 + {1 \over 6} y^3\, \k_{(2)}^2 + \cO(y^5)
\eea
For our purposes, we will need only the explicit form of the leading order coefficient,
\bea
\k_{(1)}^2 = 2 i \left(\f_-' \f_+'' - \f_-'' \f_+' \right) = - 4 \,\mathrm{Im}\left[\f_+' \bar\f_+''\right]
\eea
The function $\cG$ may be similarly expanded using (\ref{genericintro}) and the above, 
\bea
\cG = y \, \cG_{(1)} + {1 \over 6} y^3 \, \cG_{(2)} + \cO(y^5)
\eea 
with 
\bea
\label{explicitG1}
\cG_{(1)} = 4 i \left(\f_+ \f_-'- \f_- \f_+' \right) = 8 \, \mathrm{Im} \left[ \f_+ \bar \f_+'\right]
\eea
Regularity of the solutions requires that 
\bea
\label{newreqs}
\k_{(1)}^2 > 0 \hspace{1.6 in} \cG_{(1)} > 0
\eea
In fact, for generic solutions satisfying the regularity conditions (\ref{regularity1}), we may obtain explicit expressions
\bea
\label{explicitkG}
\k_{(1)}^2(x) = 2 i \sum_{k=1}^L \sum_{\ell \neq k} {Z^{[\ell k]} \over (x- r_\ell)(x-r_k)^2} \hspace{0.6 in} \cG_{(1)}(x) = 4 i \sum_{k=1}^L \sum_{\ell \neq k} {Z^{[\ell k]} \over x - r_k} \log \left|{x - r_\ell \over r_\ell - r_k}\right|\hspace{0.15 in}
\eea 
We will also need the result for the axion-dilaton $B$ on the boundary, which is found to be
\bea
\label{Bbound}
B = {2 \, \f_+ \k_{(1)} - i \,\f_+' \sqrt{6 \cG_{(1)}} \over 2 \,\bar \f_+ \k_{(1)} - i \,\bar \f_+' \sqrt{6 \cG_{(1)}}}
\eea

We may now use these results to obtain an expression for the Einstein-frame metric factor $f_6^2$, the dilaton $e^\phi$, and the string-frame metric factor $\tilde{f}_6^2$. The Einstein-frame metric factor is derived by using its form in (\ref{improvedmetricfactors}) and the above expansions. The result is surprisingly simple, 
\bea
f_6^2 = \left[{24 \, \cG_{(1)}^3 \over \k_{(1)}^2} \right]^{1/4}
\eea
The dilaton can be calculated by noting that 
\bea
e^{- 2 \phi} = \mathrm{Re} \left[{1 - B \over 1+B} \right] = {1 - B \bar B \over (1+B)(1 + \bar B)}
\eea
Then using the boundary expression for $B$, we find 
\bea
e^{-2 \phi} = {\k_{(1)} \, \cG_{(1)} \, \sqrt{6 \cG_{(1)}}\over 24 \cG_{(1)} \left(\mathrm{Re}\, \f_+' \right)^2 + 16 \k_{(1)}^2 \left(\mathrm{Re} \,\f_+ \right)^2}
\eea
The two results above may now be combined to get an expression for the string-frame metric factor $\tilde{f}_6^2$. It reduces to the following form 
\bea
\tilde{f}_6^2 = e^\phi f_6^2  = 4 \sqrt{2} \, \sqrt{ \left(\mathrm{Re}\, \f_+\right)^2 + {3 \over 2} {\cG_{(1)} \over \k_{(1)}^2} \, \left(\mathrm{Re}\, \f_+' \right)^2}
\eea
\newpage

\end{document}